\documentclass[conference]{IEEEtran}
\usepackage{cite}
\usepackage[hidelinks]{hyperref}
\hypersetup{
  colorlinks   = true, 
  urlcolor     = blue, 
  linkcolor    = blue, 
  citecolor   = blue 
}

\usepackage[ruled,linesnumbered, noend]{algorithm2e}
\usepackage{multirow}
\usepackage{tabularx,ragged2e}

\usepackage{graphicx}
\usepackage[labelformat=simple]{subcaption} 
\usepackage{threeparttable}
\usepackage{diagbox}
\usepackage{booktabs}
\usepackage{xcolor}
\usepackage{amsmath}
\usepackage{pifont}
\usepackage{numprint}
\usepackage{bm}
\usepackage{listings}
\definecolor{codegreen}{rgb}{0,0.6,0}
\definecolor{codegray}{rgb}{0.5,0.5,0.5}
\definecolor{codepurple}{rgb}{0.58,0,0.82}
\definecolor{backcolour}{rgb}{0.95,0.95,0.92}
\lstdefinestyle{mystyle}{
    language=Java,
    aboveskip=0pt,belowskip=0pt,
    commentstyle=\color{codegray},
    keywordstyle=\color{blue},
    numberstyle=\tiny\color{codegray},
    basicstyle=\footnotesize,
    xleftmargin=1em,xrightmargin=1em,
    breaklines=true,                 
    captionpos=b,                    
    keepspaces=true,                 
    numbers=left,                    
    numbersep=5pt, 
    showspaces=false,                
    showstringspaces=false,
    showtabs=false,                  
    tabsize=1
}
\usepackage{filecontents}

\usepackage{xparse}
\usepackage[scaled]{beramono}
\NewDocumentCommand{\codeword}{v}{%
\texttt{\textcolor{black}{#1}}%
}
\newcommand{\cmark}{\text{\ding{52}}}
\newcommand{\xmark}{\text{\ding{56}}}
\newcommand{\semicmark}{\cmark\kern-1.1ex\raisebox{.7ex}{\textbf{\rotatebox[origin=c]{125}{--}}}}
\newcommand{\square}{\text{\ding{110}}}

\ifCLASSINFOpdf
\else
\fi
\usepackage{url}

\begin{document}
%
\title{\textsc{BrutePrint}: Expose Smartphone Fingerprint Authentication to Brute-force Attack}

\IEEEoverridecommandlockouts
\author{
\IEEEauthorblockN{Yu Chen}
\IEEEauthorblockA{Xuanwu Lab, Tencent} \vspace{-.6cm}
\and
\IEEEauthorblockN{Yiling He}
\IEEEauthorblockA{Zhejiang University} 
\IEEEcompsocitemizethanks{\IEEEcompsocthanksitem Work done during an intership at Tencent XuanWu Lab.}
\vspace{-.6cm}
}


%


\maketitle

\begin{abstract}
Fingerprint authentication has been widely adopted on smartphones to complement traditional password authentication, making it a tempting target for attackers. 
The smartphone industry is fully aware of existing threats, and especially for the presentation attack studied by most prior works, the threats are nearly eliminated by liveness detection and attempt limit.
In this paper, we study the seemingly impossible fingerprint brute-force attack on off-the-shelf smartphones and propose a generic attack framework.
We implement \textsc{BrutePrint} to automate the attack, that acts as a middleman to bypass attempt limit and hijack fingerprint images.
Specifically, the bypassing exploits two zero-day vulnerabilities in smartphone fingerprint authentication~(SFA) framework, and the hijacking leverages the simplicity of SPI protocol.
Moreover, we consider a practical cross-device attack scenario and tackle the liveness and matching problems with neural style transfer~(NST).
A case study shows that we always bypasses liveness detection and attempt limit while 71\% spoofs are accepted.
We evaluate \textsc{BrutePrint} on 10 representative smartphones from top-5 vendors and 3 typical types of applications involving screen lock, payment, and privacy.
As all of them are vulnerable to some extent, fingerprint brute-force attack is validated on on all devices except iPhone, where the shortest time to unlock the smartphone without prior knowledge about the victim is estimated at 40~minutes.
Furthermore, we suggest software and hardware mitigation measures.
\end{abstract}

\begin{IEEEkeywords}
fingerprint authentication, brute-force attack, off-the-shelf smartphones
\end{IEEEkeywords}

\section{Introduction}

Biometric technology has advanced rapidly in recent years. Since Apple introduced Touch ID in 2013, the modern-day fingerprint sensor revolution begins, and fingerprints have become the most preferred biometric traits on smartphones~\cite{mordor_rep}. Nowadays, Smartphone Fingerprint Authentication~(SFA) is used in a variety of applications ranging from unlocking the device to authorizing payments. As a result, the security of SFA is extremely important. 

The \emph{presentation attack} has long been identified as a severe threat to the security of fingerprint authentication systems~\cite{iso2016-1}. 
The attack impersonates a target victim by presenting artefacts~(e.g., silica gel fingers) to the fingerprint sensor and has a chance of success with careful fabrication.
Prior works~\cite{plotnikov2014solution, spinoulas2021multi, kolberg2021anomaly} focus a lot on developing presentation attack detection~(PAD) algorithms, most typically for liveness detection.
However, when it comes to smartphone, such spoofing becomes much more difficult.
Firstly, some security strategies, especially the strict limit on the maximum number of failed attempts~(hereinafter ``attempt limit''), leave adversaries with rare opportunities for refining an artefact to be ambiguous for a target SFA system.
While, on the other hand, the refinement highly relies on the collection of victim-specific fingerprints and requires much expert efforts~\cite{myth_fp}.
Secondly, the commonly adopted liveness detection, that directly rejects non-live fingerprint images by material indicators such as texture and intensity~\cite{tan2006comparison}, increases the burden in artefact fabrication.
Therefore, the attack \textit{cannot be put at any scale}.

In another aspect, although several \emph{software vulnerabilities in SFA systems} have been disclosed~\cite{jo2016security, huawei_cracks}, they \textit{only affect specific smartphone models and operating system~(OS) versions}.
Those newly patched vulnerabilities can be divided into two categories.
(a)~Unwary fingerprint authentication algorithms: some textures on top of fingerprint scanners interfere with the algorithms~(e.g., screen protector problems in Samsung~\cite{CVE-2019-17668, CVE-2021-22494}), making a zero-effort attempt possible to match. Nevertheless, these vulnerabilities are accidentally exploitable when the textures trigger fingerprint template update in the algorithm.
(b)~Improper implementations with software: these vulnerabilities are not actually exploitable without certain permissions~\cite{CVE-2020-11600}.
For instance, Leftover Debug Code found on OnePlus 7 Pro before 10.0.3.GM21BA~\cite{CVE-2020-7958} allows fingerprint image interception only when the attacker gains root access and the victim cooperates in finger-pressing without consciousness.

Given these situations, is authenticating with fingerprints secure enough on our smartphones? This work gives a brand new threat that \textit{large-scale} fingerprint brute-force attack is practical on \textit{off-the-shelf} smartphones.
We discover Cancel-After-Match-Fail~(CAMF) and Match-After-Lock~(MAL) vulnerabilities in SFA, and either of them can be exploited to bypass attempt limit.
Instead of an implementation bug, CAMF and MAL leverage logical defects in the authentication framework.
Therefore, it \textit{exists across various models and OSs}.
We make the exploitation on 10 popular smartphone models with the latest OS versions.
Results show that attempts are made three times over the attempt limit on Touch ID while unlimited attempts are achieved on all Android devices.
The unlimited attempts motivate us to perform automatic fingerprint brute-force attacks. 

We find the insufficient protection of fingerprint data on the Serial Peripheral Interface~(SPI) of fingerprint sensors, and thus come up with a hardware approach to do man-in-the-middle~(MITM) attacks for fingerprint image hijacking. We design the adversarial equipment to be compatible with different smartphone models and show an implementation that is low-cost and novice-friendly. 
We propose using neural style transfer~(NST) techniques to make arbitrary outer fingerprint image database adoptable for brute-forcing.
Experiments show that the proposed method bypasses liveness detection as well as reserving enough fingerprint features to pass fingerprint matching. 
We check the feasibility of fingerprint image hijacking and the practicality of fingerprint brute-force attacks on the 10 smartphones.
The affected smartphones cover fingerprint sensors of commercially available types, i.e., optical~(ultra-thin included), capacitive, and ultrasonic.
To illustrate, we show that our attacks can be used to unlock screen locks, authorize payments, and unlock privacy apps.

\noindent\textbf{Contributions.} This paper has three major contributions:
\begin{itemize}
    \item We discover software and hardware vulnerabilities on off-the-shelf smartphone fingerprint authentication systems. We exploit the vulnerabilities to bypass the attempt limit in OS and hijack fingerprint images on SPI. We propose fingerprint brute-force attack on smartphones and leverage neural style transfer to ease the attack.
    \item We validate attempt limit bypassing and fingerprint image hijacking on 10 prevalent smartphone models with various OS versions and fingerprint sensors. For bypassing, we achieve infinite attempts on Android/HarmonyOS devices while making 10 additional attempts on iOS devices. For hijacking, fingerprint image interception and replacement are achieved on all devices except iPhone.
    \item We implement \textsc{BrutePrint} with low-cost equipment and show a complete fingerprint brute-force attack process on a OnePlus 7 Pro. We demonstrate how the attack is practical on other devices and confirm that 3 typical types of fingerprint applications are affected. 
\end{itemize}


\section{Background}\label{sec:background}

In this section, we introduce background on current SFA systems concerning authentication process, security enhancement, and evaluation metrics.

\begin{figure}[htbp]
    \centering
    \includegraphics[width=1\linewidth]{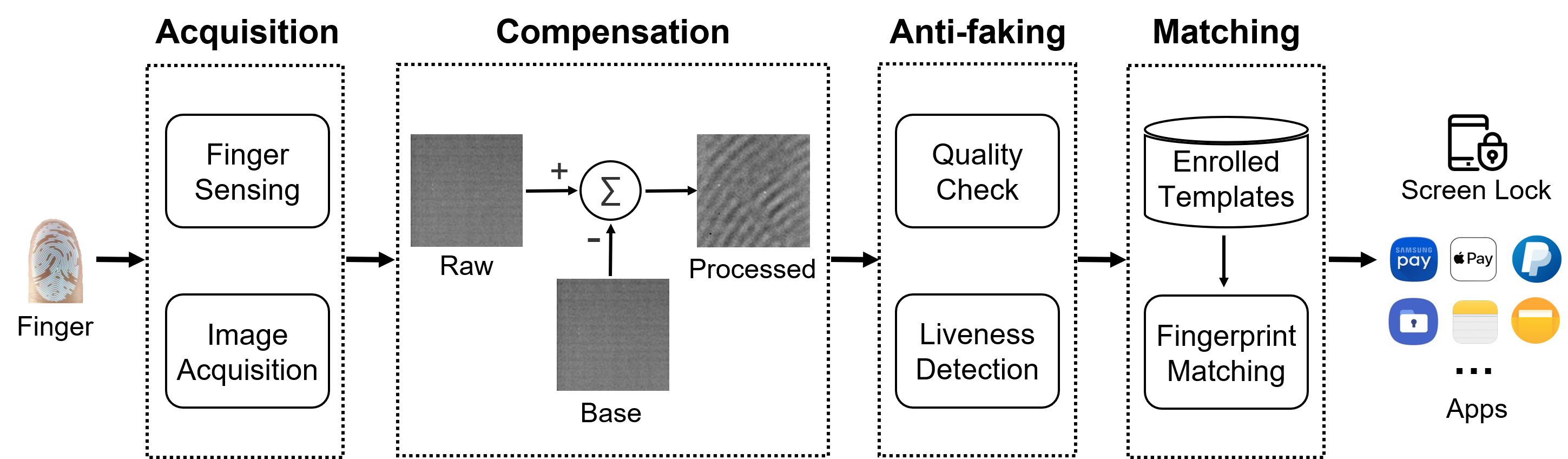}
    \caption{The workflow of state-of-the-art SFA systems.}
    \label{fig:sfa}
\end{figure}

\subsection{Authentication Process}
Fingerprint authentication process on smartphones consists of four main stages, namely \emph{Acquisition}, \emph{Compensation}, \emph{Anti-faking}, \emph{Matching}, as illustrated in Figure~\ref{fig:sfa}. 

(a)~Acquisition starts fingerprint sensing when SFA systems receive fingerprint authentication requests. If a finger-pressing is sensed, the image acquisition module captures one or multiple raw fingerprint images. 

(b)~Compensation processes the raw fingerprint images to boost image quality, where some fixed textures are reduced through a pre-calculated base image.
The base image contains fixed pattern noise~(FPN) caused by screen pixels, screen protectors, charge transfer, etc, and the processed fingerprint image is gained through an additive operation.

(c)~Anti-faking checks the quality of processed images and applies liveness detection in order to prevent presentation attacks. 
SFA systems integrate the liveness detection through image analysis.
As fake fingers are made with fabrication materials and lead to different behaviors of finger-pressing, the methods are intuitive that identify outliers with features including material textures and ridge-valley contrast~\cite{tan2010spoofing}.

(d)~Matching is to measure the similarity between query and enrolled fingerprints. 
To minimize the damage of fingerprint leakage, the enrolled fingerprints are stored as irreversible templates~(e.g., mathematical representations) rather than fingerprint images. Thus, the similarity calculation is done on the processed fingerprint images and the enrolled fingerprint templates.
To maintain the security of small mobile sensors and improve user experience~\cite{park2008fingerprint}, fingerprint matching algorithms in SFA systems also evolve significantly to handle partial fingerprint images.

\subsection{Security Enhancement}
Besides the anti-faking, security enhancement in SFA incorporates risk control strategies. The strategies prohibit fingerprint authentication and require passcode under exceptional circumstances. Three typical examples are that (a)~the number of unrecognized fingerprint attempts in a row exceeds the attempt limit; (b)~the time passed from the last device unlocking is longer than the idle timeout; (c)~the smartphone restarts.

Trusted Execution Environment (TEE) is used to isolate fingerprint sensor driver, fingerprint authentication logic, and fingerprint data in a secure environment. With this design, the fingerprint authentication process stays secure even if Rich Execution Environment~(REE) is compromised.
In the case of smartphones, the REE runs OSs such as Android and iOS. 
Since Android 6, Google enforces moving all fingerprint data manipulation into TEE~\cite{android6}, and most smartphone manufacturers adopt TrustZone-based TEE solutions.
At nearly the same time, Apple uses another solution called Secure Enclave~\cite{touchid2017}.

\subsection{Evaluation Metrics}
The error rate in a SFA system is measured with two metrics~\cite{iso2006-1}. 
(a) False Accept Rate~(FAR): proportion of authentication transactions with wrongful claims of identity that are incorrectly confirmed.
(b)~False Reject Rate~(FRR): proportion of authentication transactions with truthful claims of identity that are incorrectly denied.

FRR at fixed levels of FAR are used to determine if a system has sufficient accuracy for a specific use case.~\cite{biometrics2014understanding}
To ensure the security, SFA systems should have a FAR not higher than $1/50000$~\cite{android6}.
Computing FAR uses collected fingerprint databases, where multiple fingerprint images are collected from each subject to make the collection efficient. A common practice is to set the minimum subject number, the finger number per subject and the sample image number per finger to $24$, $4$~(left thumb, left index, right thumb, right index) and $50$, respectively. 
This setting makes $(24\times4-1)\times50\times(24\times4)=456000$ wrongful claims for the FAR evaluation, and a qualified system should give incorrect confirmations $9$ times at most. 

\section{Attack Framework}\label{sec:framework}

\subsection{Threat Model}
We consider fingerprint brute-force attack on smartphones. 
The \emph{goal} is to make unlimited submission of fingerprint images and eventually guess a fingerprint correctly.
Unlike traditional password authentication, fingerprint authentication decides a match with a reference threshold instead of a specific value.
Therefore, fingerprint brute-force attack can succeed by leveraging the FAR of a target system.
The success rate at time $t$ can be estimated as
\begin{equation}
    1-\left(1-r \cdot \mathrm{FAR}\right)^{\bm{\mathrm{FIPS}} \cdot t}\,
\label{equ:suc_rate}
\end{equation}
where $r$ is the number of enrolled fingerprints on a victim device, and $\bm{\mathrm{FIPS}}$ defines the number of submitted fingerprint images per second.
In theory, the attack is successful if two requirements are satisfied: the number of attempts and the denominator of the FAR are in the same order of magnitude~(\emph{R1}); the submitted fingerprints are totally controllable~(\emph{R2}).
Two main \emph{threats} of a successful attack are that attackers can get (similar) fingerprints of the victim~(\emph{T1}) and gain authorization in many apps on the victim device~(\emph{T2}). 

\textbf{Assumption.} We assume that the targeted smartphone supports fingerprint authentication.
Adversaries can have physical access to the victim device, but they cannot root it and know nothing about the owner. 
To launch the attack, they need to own a fingerprint database and an adversarial equipment that hijacks data sent by fingerprint sensor.
\begin{itemize}
    \item \emph{Physical Access.} Adversaries can remove the rear cover of the smartphone and place the adversarial equipment for several hours. This would be possible when the victim device is lost, stolen, temporarily deposited, unattended during owner's sleep, etc.
    \item \emph{Adversarial Equipment.} The adversarial equipment is mainly a printed circuit board~(PCB), which is inexpensive and universal. For specific smartphone models, adaptive flexible printed circuit~(FPC) is required. The equipment costs around 15 dollars in total.
    \item \emph{Fingerprint Database.} The database can originate from any sources and have fingerprint images captured by arbitrary sensors. The images can be gathered from academic dataset~\cite{shehu2018sokoto, sd301}, data breaches~\cite{biostar_leak}, and self-collection. It's possible for adversaries to have a database containing fingerprints of millions of identities~\cite{brazil_fp_leak}.
\end{itemize}

\subsection{Attack Overview}

\begin{figure}[htbp]
    \centering
    \includegraphics[width=1\linewidth]{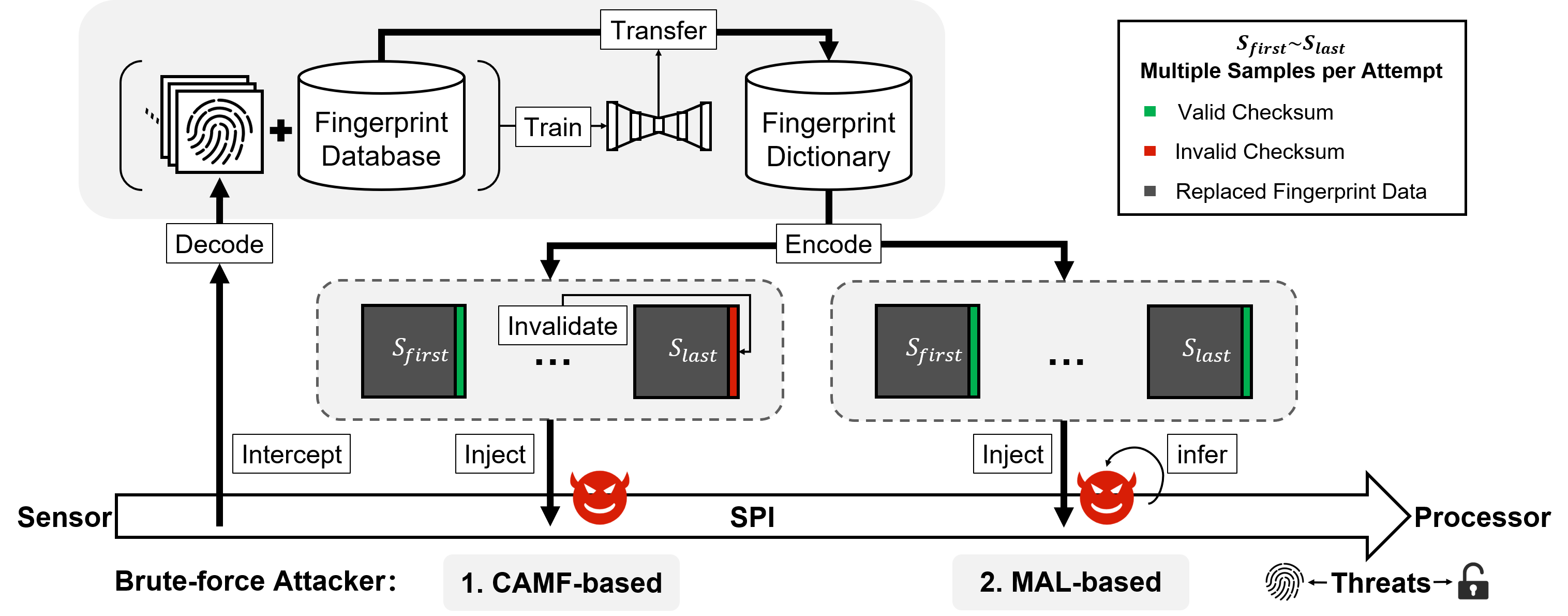}
    \caption{The attack overview of \textsc{BrutePrint}.}
    \label{fig:overview}
\end{figure}

We implement \textsc{BrutePrint} system to crack SFA and achieve fingerprint brute-force attack.
As shown in Figure~\ref{fig:overview}, there are two kinds of brute-force attackers who exploit CAMF and MAL vulnerabilities. 
Specifically, \textsc{BrutePrint} acts as a middleman between fingerprint sensor and TEE.
Two attacks are firstly implemented to reach ultimate fingerprint brute-force attack. 
\textbf{Attempt limit bypassing} attack exploits either of the two vulnerabilities to meet R1.
Typically, a CAMF exploitation \emph{invalidate}s the checksum of transmitted fingerprint data, and a MAL exploitation \emph{infer}s matching results through side-channel attacks.
\textbf{Fingerprint image hijacking} attack meets R2, which has the capability to \emph{decode} the \emph{intercept}ed fingerprint data and \emph{encode} replaced data for \emph{inject}ion.
To increase the success rate of brute-forcing, \textsc{BrutePrint} additionally propose a \textbf{fingerprint dictionary generation} method that \emph{train}s a neural style transfer network to \emph{transfer} available fingerprint database into valid styles.

\section{Methodology}\label{sec:method}

\subsection{Attempt Limit Bypassing}

\subsubsection{CAMF Vulnerability}
CAMF is based on the fault-tolerant mechanisms in SFA systems, where collecting multiple samples in a single authentication attempt is considered as one of the best practices~\cite{mansfield2002best}. 
Two key points of the mechanisms are related with the vulnerability: 
\begin{itemize}
    \item \emph{Multi-sampling.} To tolerate the false rejection of fingerprint matching algorithms, an attempt can pass the authentication if one sample passes.
    \item \emph{Error-cancel.} To tolerate some minor errors (e.g., caused by a temporary hardware malfunction), a failed attempt will be canceled if certain error signals are received.
\end{itemize}
\begin{figure}[htbp]
    \centering
    \includegraphics[width=\linewidth]{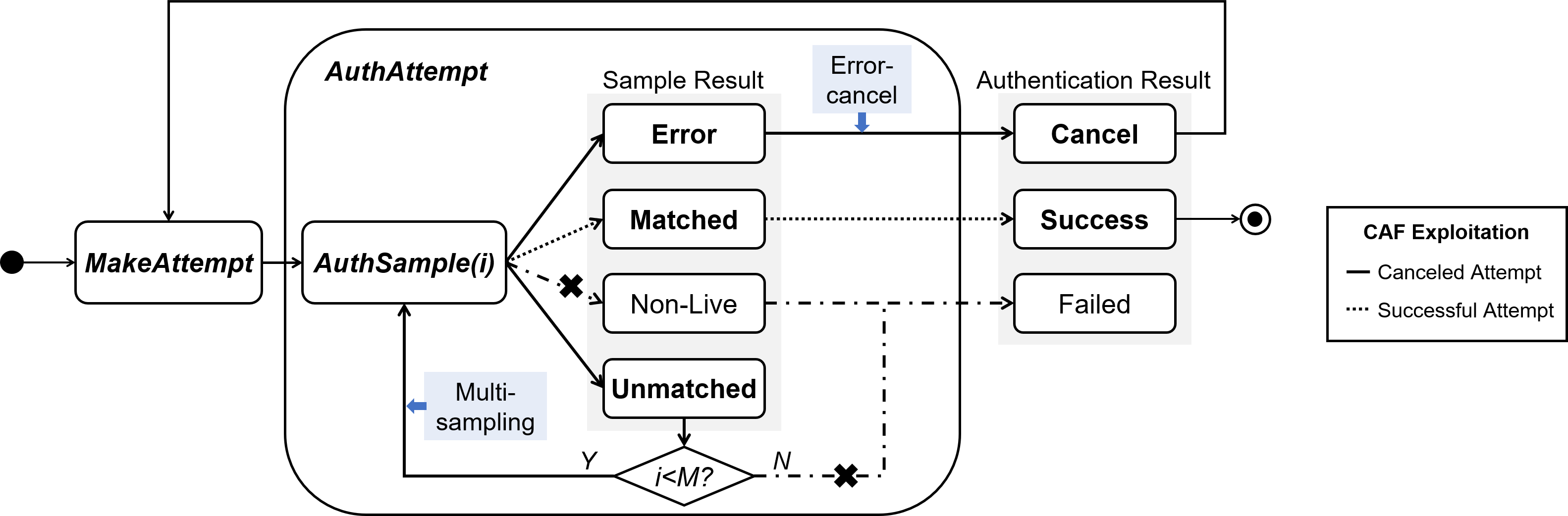}
    \caption{Explanation of CAMF vulnerability and exploitation.}
    \label{fig:caf-logic}
\end{figure}
As shown in Figure~\ref{fig:caf-logic}, the authentication for an attempt acquires and authenticates samples in a loop until a sample result is \textit{Error/Success/Non-live} or the sample number reaches the maximum $M$.
Only a \textit{Failed} authentication result decreases the remaining attempt number restricted by attempt limit. 

Therefore, attackers can make unlimited attempts if the last, i.e., $M$-th, sample of each attempt goes into the Error-cancel branch.
The attempt is either successful or canceled, depending on whether a matched sample is in the first $(M-1)$ samples.
The \textit{Failed} authentication result is avoided through cancelling after $(M-1)$ unmatched sample results.

\subsubsection{MAL Vulnerability}
MAL refers to the vulnerability that attackers can make attempts to infer authentication results of fingerprint images~(called the ``inference attempt'') in lockout mode. 
The lockout mode is defined in Google's biometric framework to penalize too many failed attempts, where no fingerprint authentication can be launched in a certain period of time or permanently. 
Listing~\ref{lst:mal} shows how the mode affects the authentication.
A finger-pressing is not supposed to get any response in \texttt{lockoutMode} since the method \texttt{startAuthentication} is directly returned without \texttt{startClient}.
However, a pseudo lockout state is introduced in some smartphone models in order to improve user experience when waking up the locked screen.
The highlighted code snippet is the modification made by a smartphone manufacture, where lockout mode is ignored for \texttt{Keyguard} process that handles the screen lock.

\vspace{1mm}
\fbox{
  \begin{minipage}[htbp]{.45\textwidth}
    \centering
    \lstinputlisting[lastline=5]{mal.java}
    \lstinputlisting[backgroundcolor=\color{yellow},firstline=6,lastline=6, firstnumber=6]{mal.java}
    \lstinputlisting[firstline=7, firstnumber=7]{mal.java}
  \end{minipage}
}
\lstinputlisting[caption=Explanation of MAL vulnerability, label={lst:mal}]{blank.java}

\noindent Although the lockout mode is further checked in \texttt{Keyguard} to disable unlocking, the authentication result has been made by TEE. As \textit{Success} authentication result is immediately returned when a matched sample is met, it's possible for side-channel attacks to infer the result from behaviors such as response time and the number of acquired images. 

The pseudo lockout mode as well as the side-channel information enable MAL-based attackers to make unlimited submission of fingerprint samples to find a match. 
Also, the matched fingerprint can be reused in a normal mode to unlock the device. 
Compared with CAMF, MAL only exist in the screen lock application and may has extra time cost~(see Section~\ref{sec:rate}), but it have an advantage that the exploitation can be made even if fingerprint authentication is permanently locked on the victim device~(e.g., after the idle timeout or being restarted).

\subsection{Fingerprint Image Hijacking} \label{sec:hijack_method}

\textbf{MITM Attack on SPI.}
In SFA systems, Serial Peripheral Interface~(SPI) is used for synchronous serial communication between fingerprint sensor and smartphone processor.
Specifically, the SPI is a four-wire bus consisting of master input/slave output~(MISO), master output/slave input~(MOSI), a serial clock~(SCLK), and a chip select~(CS) pin, where processor is the single master and sensor is the single slave.
Typically, the SCLK has a medium frequency range of 8\textasciitilde50~MHz, and thus checksum is required for resisting external disturbance. However, to the best of our knowledge, SFA sensors except Touch ID do not encrypt any data and lack mutual authentication. Together with the frequency that is possible for injection, the situation leads SFA vulnerable to MITM attack on SPI.

\begin{figure}[htbp]
\vspace{-0.3cm}
    \centering
    \includegraphics[width=.85\linewidth]{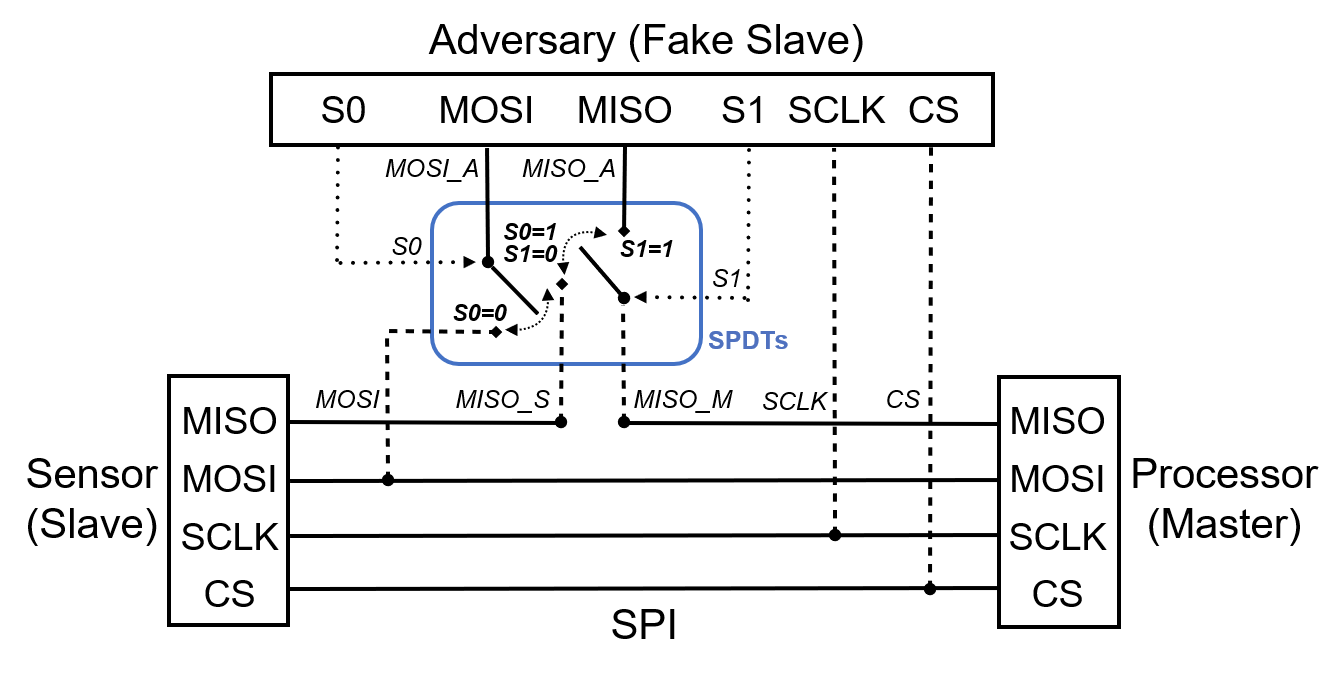}
    \caption{MITM Attack on SPI. Different status of SPDTs are explained in Table~\ref{tab:SPDT}.}
    \label{fig:mitm}
\vspace{-0.5cm}
\end{figure}

\textsc{BrutePrint} designs the MITM attack as in Figure~\ref{fig:mitm}.
The adversary acts as a fake slave to receive \emph{MOSI\_A} signal and send \emph{MISO\_A} signal, which is controlled by two Single Pole Double Throw (SPDT) switches.
The initial state for switch \emph{S0} and \emph{S1} is $(0, 0)$ that listens for fingerprint data acquisition~(FDA) for specific attacks.
Based on this implementation, possible attacks includes (a)~\textbf{glitch attack:} randomly change the state to $(0,1)$ for an instant during fingerprint data transmission to inject a glitch; (b)~\textbf{fingerprint image eavesdropping:} intercept fingerprint data with the state $(1,0)$ and decode it to get a raw fingerprint image; (c)~\textbf{fingerprint image replacement:} with the state $(1,1)$, intercept the original data and inject a replaced raw fingerprint image that is correctly encoded.

\begin{table}[htbp]
\vspace{-0.2cm}
\setlength{\abovecaptionskip}{0.cm}
\setlength{\belowcaptionskip}{0.1cm}
\centering
\caption{Function of the two SPDT switches.}
\begin{tabular}{>{\centering\arraybackslash}p{0.058\linewidth}%
		>{\centering\arraybackslash}p{0.058\linewidth}%
		>{\centering\arraybackslash}p{0.74\linewidth}}
\toprule
Switch & Status & Function                                                          \\ \midrule
\multirow{2}{*}{\emph{S0}} & 0 & Identify the FDA command from the \emph{MOSI} signal. \\
       & 1     & \textbf{Intercept} fingerprint data from the \emph{MISO\_S} signal. \\ \midrule
\multirow{2}{*}{\emph{S1}} & 0 & Keep the original connection from \emph{MISO\_S} to \emph{MISO\_M}.                           \\
       & 1     & \textbf{Inject} fake fingerprint data from \emph{MISO\_A} to \emph{MISO\_M}. \\ \bottomrule
\end{tabular}\label{tab:SPDT}
\end{table}


\textbf{Raw Image Decoding and Encoding}
Fingerprint image hijacking is implemented to achieve both fingerprint image eavesdropping and replacement. Besides the MITM attack framework, functions for decoding and encoding raw fingerprint images are specialized for smartphone models. 
The two functions are inverses and fit the reverse engineered fingerprint data transmission protocol to deal with image shape, pixel representation, byte sequence, checksum, and frame structure.
For instance, necessary steps for encoding include pixel adaptation to unify bits per pixel~(bpp), byte reordering to compress adjacent pixels, checksum attachment to verify image data, and frame alignment to separate and add header to frames.

\subsection{Fingerprint Dictionary Generation}
Fingerprint brute-force attack can only be launched with inputs acceptable to the SFA system.
Raw fingerprint images acquired by SFA sensors have a finite resolution and non-negligible FPN~(see Figure~\ref{fig:hijack_images}). 
Therefore, suppose an attacker gets a fingerprint image database from open sources such as FVC~\cite{FVC}, he should transfer the image style into the sensor-specific style to generate a fingerprint dictionary for brute force.
We use neural style transfer techniques to accomplish two purposes: (a)~the authentication will not be terminated ahead by image-level liveness detection, i.e., not go into the non-live branch in Figure~\ref{fig:caf-logic}; (b)~the transferred images are able to remain enough fingerprint features for matching algorithms and pass the authentication.

\textbf{Style Alignment.} The problem is formalized as learning the mapping $G:X\rightarrow Y$ that translates the source domain $X$ of open source fingerprint images to the target domain $Y$ of the sensor-captured images that are specific to a smartphone model. We address three main issues as follows:
\begin{itemize}
    \item \emph{Non-uniform sources.} We unify the images from $X$ to eliminate the differences among source devices, especially to align the period of a ridge/valley cycle with the targets. The difference is brought by the Dots per Inch~(DPI) of devices, so we can scale each source image with the DPI ratio of source device to target device. In addition, we remove unnecessary space from around images and crop the central area to get the pre-processed images that have congruent shape with the targets.
    \item \emph{Unpaired translation.} Attackers can acquire the target domain images themselves by using the hijacking equipment, and thus the training data from $X$ and $Y$ is unpaired. We adopt the idea from CycleGAN~\cite{zhu2017unpaired} that addresses the highly under-constraint training problem with a coupled inverse mapping.
    \item \emph{Inflexible targets.} Since SFA have compensation module to remove FPN, any delicate alteration may trigger a rejection decision. We must be cautious that the target images for training should stay unchanged. That is to say, we cannot perform image transformations on them, e.g., cropping, resizing and flipping. As the generator of CycleGAN accepts only square inputs like most models, we adapts the stride and padding of some convolutional layers to meet the rectangle shape for $Y$.
\end{itemize}
Fig.~\ref{fig:transfer} illustrates how source images are unified and translated.
To make the training process easier, target images are sometimes refined to get the pure fingerprint area, which we will discuss in the next section.

\section{Brute-force Attack: A Case Study}\label{sec:case}

In this section, we show the whole process of brute-forcing a OnePlus 7 Pro~(called the ``victim device'' throughout the section) and give experimental results.
We explain choosing the victim device that (a)~it runs close to the stock version of Android OS, which is open-source and the most popular, making it representative for studying SFA; (b)~the embedded fingerprint sensor uses the advanced optical in-display technology that has become a built-in feature of many mid-end and high-end Android devices.
As ScreenLock is where fingerprint authentication is the most frequently used and protects other applications, we do experiments on it to achieve the harmful brute-forcing effect of screen unlocking. 
 
\subsection{Unlimited Attempt Limit Bypassing}\label{7P_bypass}
We find CAMF vulnerability on the victim device and make exploitation with checksum error-cancel.
Results show that unlimited attempts can be made with the exploitation.
We further dig into the code flow and reason the unlimited attempt limit bypassing as follows.

\begin{figure*}[htbp]
    \centering
    \begin{subfigure}[b]{.45\linewidth}
      \centering
      \includegraphics[width=\linewidth]{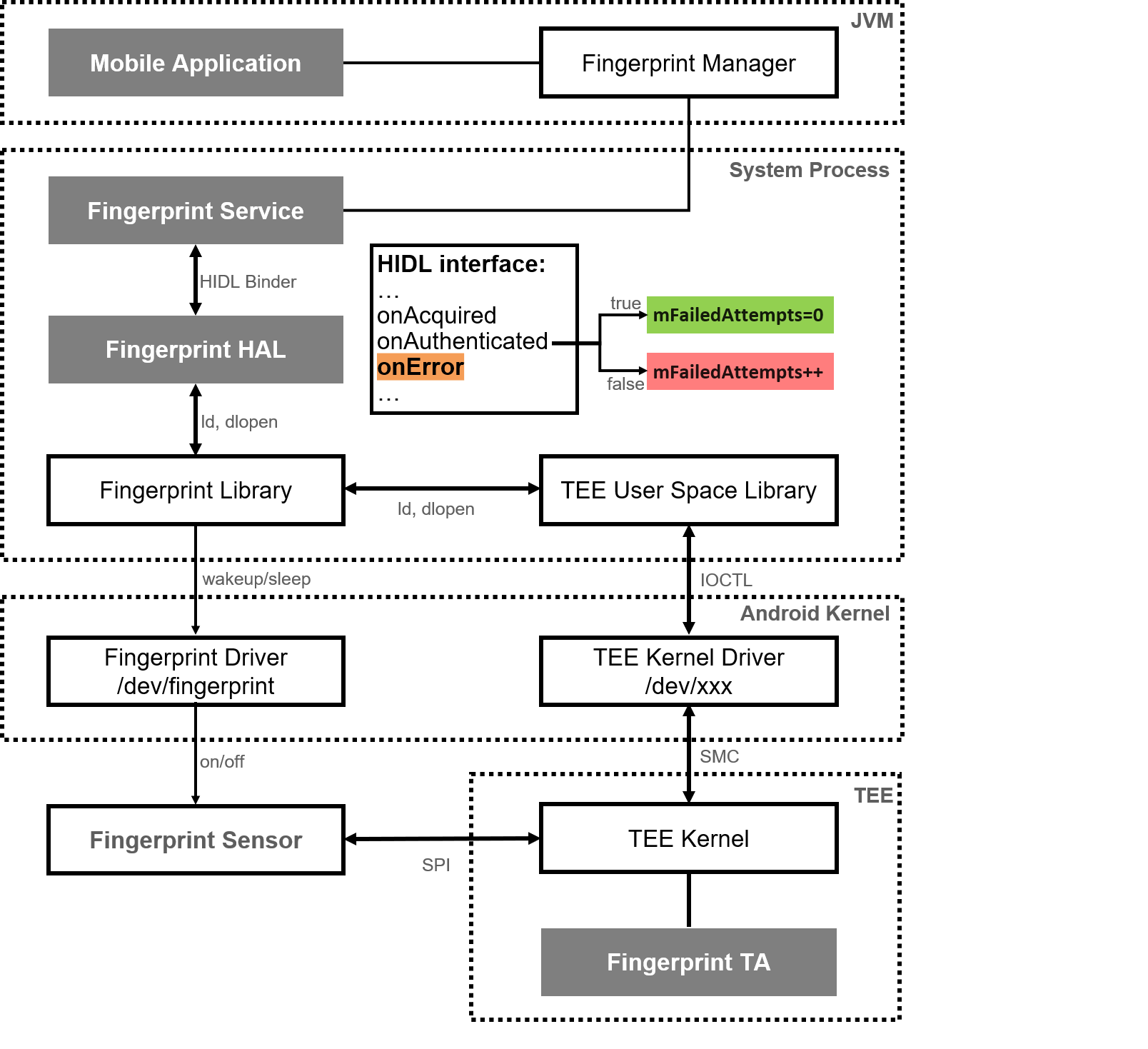}  
      \caption{SFA framework in Android.}
      \label{fig:android_fp_arch}
    \end{subfigure}
    \begin{subfigure}[b]{.45\linewidth}
      \centering
      \includegraphics[width=\linewidth]{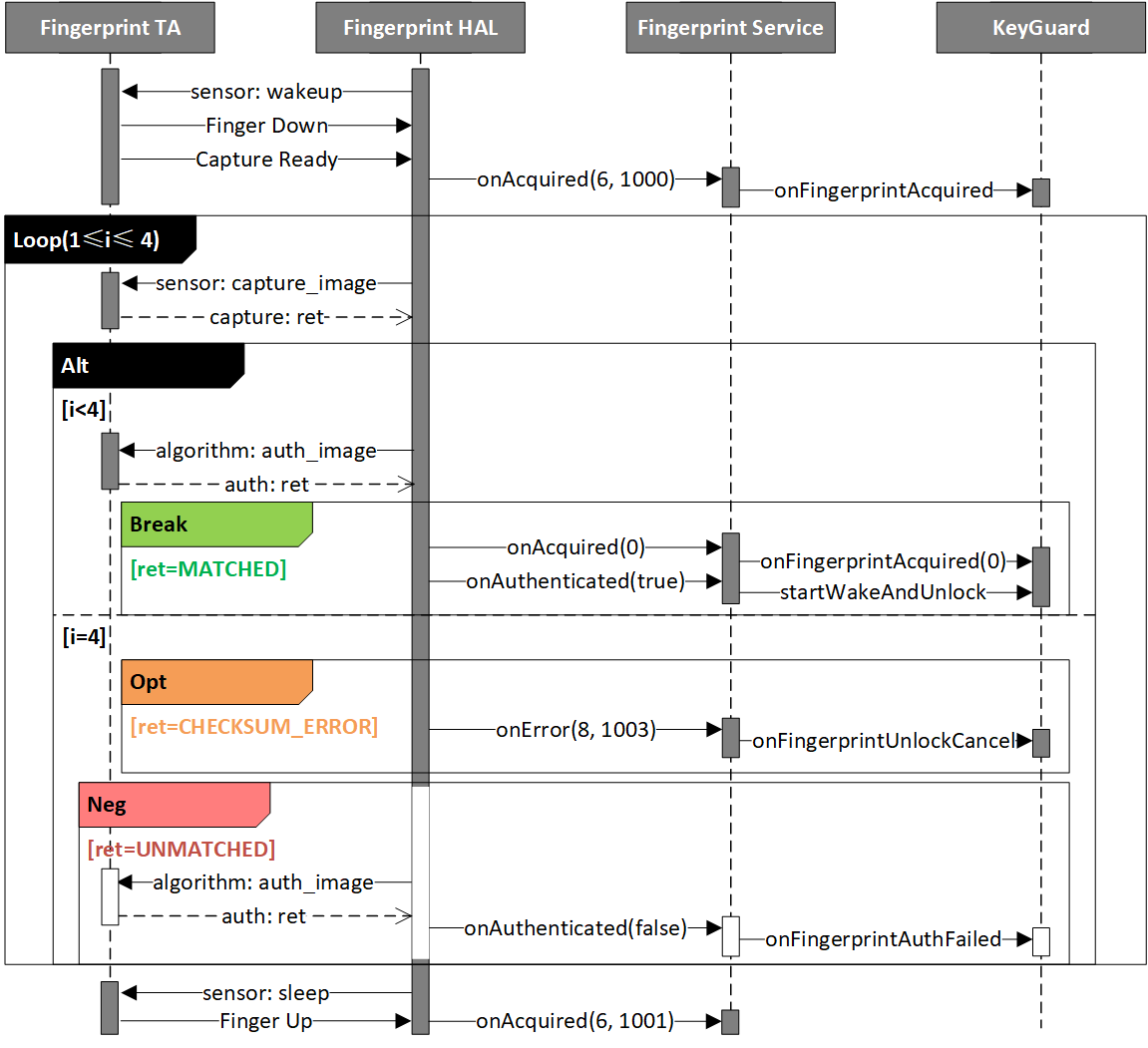}  
      \caption{Sequence diagram for attempt limit bypassing.}
      \label{fig:uml_seq_bypassing}
    \end{subfigure}
    \caption{Explanation of attempt limit bypassing attack in the case study. In Android biometric framework, fingerprint authentication is closely related to four roles, i.e., fingerprint TA, fingerprint HAL, fingerprint service, and mobile application. The interactions between them under our attack is represented with a sequence diagram.}
    \label{fig:arch}
\end{figure*}

\textbf{Android SFA Framework.}
The victim device uses Android OS that has a fingerprint authentication framework with four layers as in Figure~\ref{fig:android_fp_arch}. Each mobile application has its own manager for fingerprint authentication to communicate with the fingerprint service in system process.
The service is implemented in the class \texttt{FingerprintService} that belongs to Android Open Source Project~(AOSP) code.
It connects to fingerprint Hardware Abstraction Layer~(HAL) with fingerprint Hardware Interface Definition Language~(HIDL) to send \emph{command} to vendor-specific libraries and fingerprint sensor. 
Vendors protect their core fingerprint authentication functions in TEE, where TEE kernel communicates with fingerprint sensor through SPI and fingerprint Trusted Application~(TA) is responsible for algorithms including compensation, anti-faking, and matching.
Three important \emph{callback}s are originally defined by the framework in HIDL interface to return asynchronous results from TEE in response to user actions on fingerprint sensor:
\begin{itemize}
    \item \texttt{onAcquired}. sent when a fingerprint image is acquired by the sensor and notifies a \texttt{acquiredInfo} message about the image quality; a vendor-specific message \texttt{vendorCode} is validate when \texttt{acquiredInfo} is set to 6; 
    \item \texttt{onAuthenticated}. sent when a fingerprint is authenticated and notifies an authentication result;
    \item \texttt{onError}: sent when a fingerprint error occurs and notifies a \texttt{error} message; a vendor-specific message \texttt{vendorCode} is validate when \texttt{error} is set to 8.
\end{itemize}
The behavior of the callbacks are implemented in \texttt{FingerprintService}, where attempt limit is monitored by the variable \texttt{mFailedAttempt} and can only be modified when \texttt{onAuthenticated} is called back.
This will happen when fingerprint TA sends an \emph{event} for HAL to notify authentication results.
If the authentication result is \texttt{false}, the \texttt{mFailedAttempt} will be increased by one, which decreases the remaining number of available attempts~(i.e., attempt limit - \texttt{mFailedAttempt}). 
Therefore, a successful attack should ensure not calling \texttt{onAuthenticated} back. An alternative is to trigger an event relevant to the callback \texttt{onError}.

\textbf{Checksum Error-cancel.}
As checksum is highly required by SPI transmission protocols and is initially attached to detect accidental changes in raw fingerprint data, it could be natural for vendors to regard a checksum error event as a cancelable state for authentication. 
To bypass attempt limit for ScreenLock on the victim device, we invalidate the checksum of the fourth sample's fingerprint data and successfully trigger Error-cancel. 
To reason the exploitation, we extract critical control flow in Android SFA framework through \texttt{Logcat} and show it in Figure~\ref{fig:uml_seq_bypassing}.
Once fingerprint HAL receives an authentication request, it will send command to wake up sensor and notify the \texttt{onAcquired} callback. 
Then the authentication loop goes through image capturing and image authentication for each sample until the last one, which notifies the \texttt{onError} callback with a vendor-specific message after image capturing. 
It's worth mentioning that under our attack, the loop never ends with a unmatched sample result to notify \texttt{onAuthenticated} with a failed authentication result.
Moreover, if a sample is matched in the first three authentication process, the HAL will notify \texttt{onAcquired} with \texttt{FINGERPRINT\_ACQUIRED\_GOOD} message, send a success authentication result to \texttt{onAuthenticated}, and directly break the loop.

\subsection{Successful Fingerprint Image Hijacking}\label{case:hijacking}
Fingerprint image hijacking observes the sensor's SPI protocol for fingerprint data.
We find the data not encrypted and successfully infer the FDA command as well as how the fingerprint image is encoded.
We make a hardware implementation of the SPI MITM and use it to locate fingerprint data, recover fingerprint image from the data, and replace the original data with the re-encoded image. Experiments show that authentication results are returned without influence, that is, the replaced data of a registered finger can still unlock the victim device.
Next, we describe the detailed hardware implementation and the reverse engineered SPI protocol.

\textbf{Implementation of SPI MITM.}
The hardware implementation of Figure~\ref{fig:mitm} requires a fake SPI slave and the SPDTs.
There are plenty of approaches to fake the slave, but we carefully consider the price and choose to use the SPI peripheral of a MCU with slave configuration.
Moreover, it should satisfy the 25~MHz fingerprint transmission rate on the victim device.
For the SPDTs, the switching time should be less than the 206~$\mu$s interval between identifying FDA on MOSI and the beginning of fingerprint data transmission on MISO.
Therefore, we choose a high bandwidth dual-channel SPDT chip.

\begin{figure}[htbp]
    \centering
    \includegraphics[width=\linewidth]{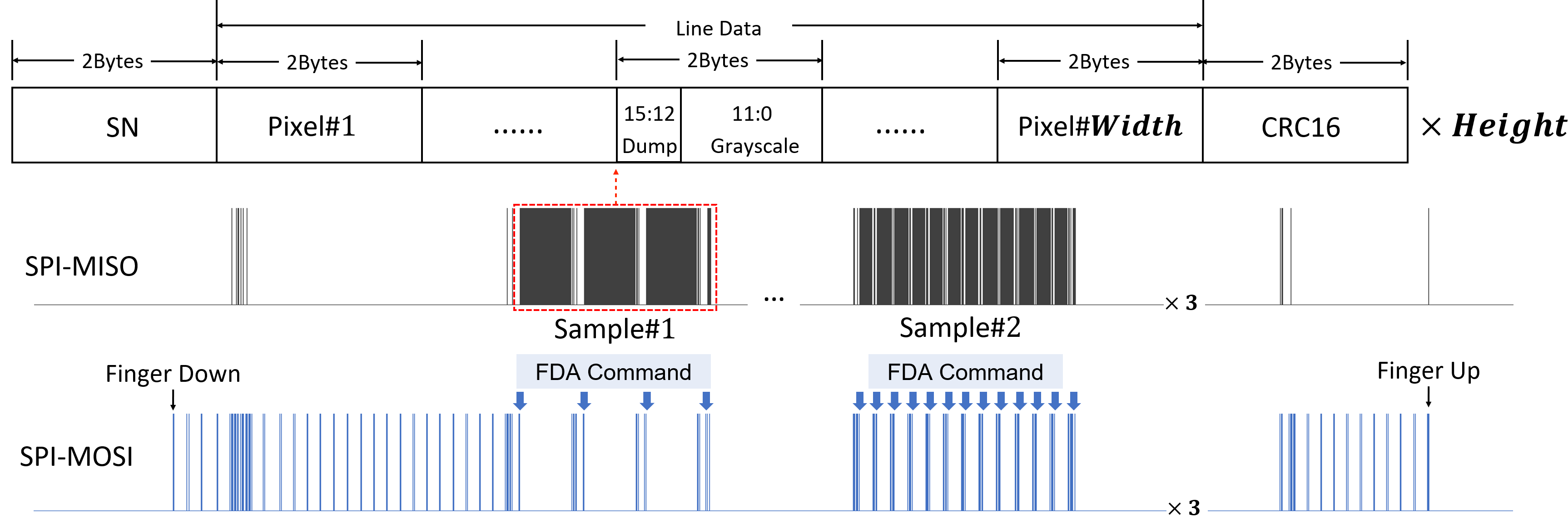}
    \caption{Example of reverse engineered SPI protocol for fingerprint data on MISO and MOSI.}
    \label{fig:protocol}
\vspace{-2mm}
\end{figure}

\textbf{Reverse Engineered Protocol.}
We capture SPI signals through a logic analyzer and locate those dense signals on MISO to identify FDA. As the data is not encrypted, we could try out the encoding method within several inference attempts. For example, the image shape can be guessed through factoring the total number of pixels, and adaption can be made according to the periodic offset of outlier values~(i.e., values such as checksum other than image pixels). 
For the victim device, the first sample is transmitted in 4 frames while the last three use the same format with 13 frames. Each last frame is short since it transmits the remained fingerprint data.
The FDA commands are identified before every frame.
Taking the first sample for example, the FDA commands are always \texttt{0xF08800}, and we show the structure~(frame separator omitted) of fingerprint data in Figure~\ref{fig:protocol}. The structure is not complicated that the gray-scale image is stored in 16~bpp, and for each line, a serial number and a CRC16 checksum are attached at both ends.

\subsection{Effective Dictionary Generation for Brute-force}\label{sec:dic_case}

Based on the aforementioned CAMF exploitation and hijacking equipment, we can perform fingerprint brute-force attack on the victim device.
Considering a weak assumption of attack scenario, that adversaries own a fingerprint database acquired and processed by sensors totally different from the victim device's sensor, we stimulate the scenario and evaluate the fingerprint dictionary generation method as follows.

\begin{figure}[htbp]
    \centering
    \includegraphics[width=\linewidth]{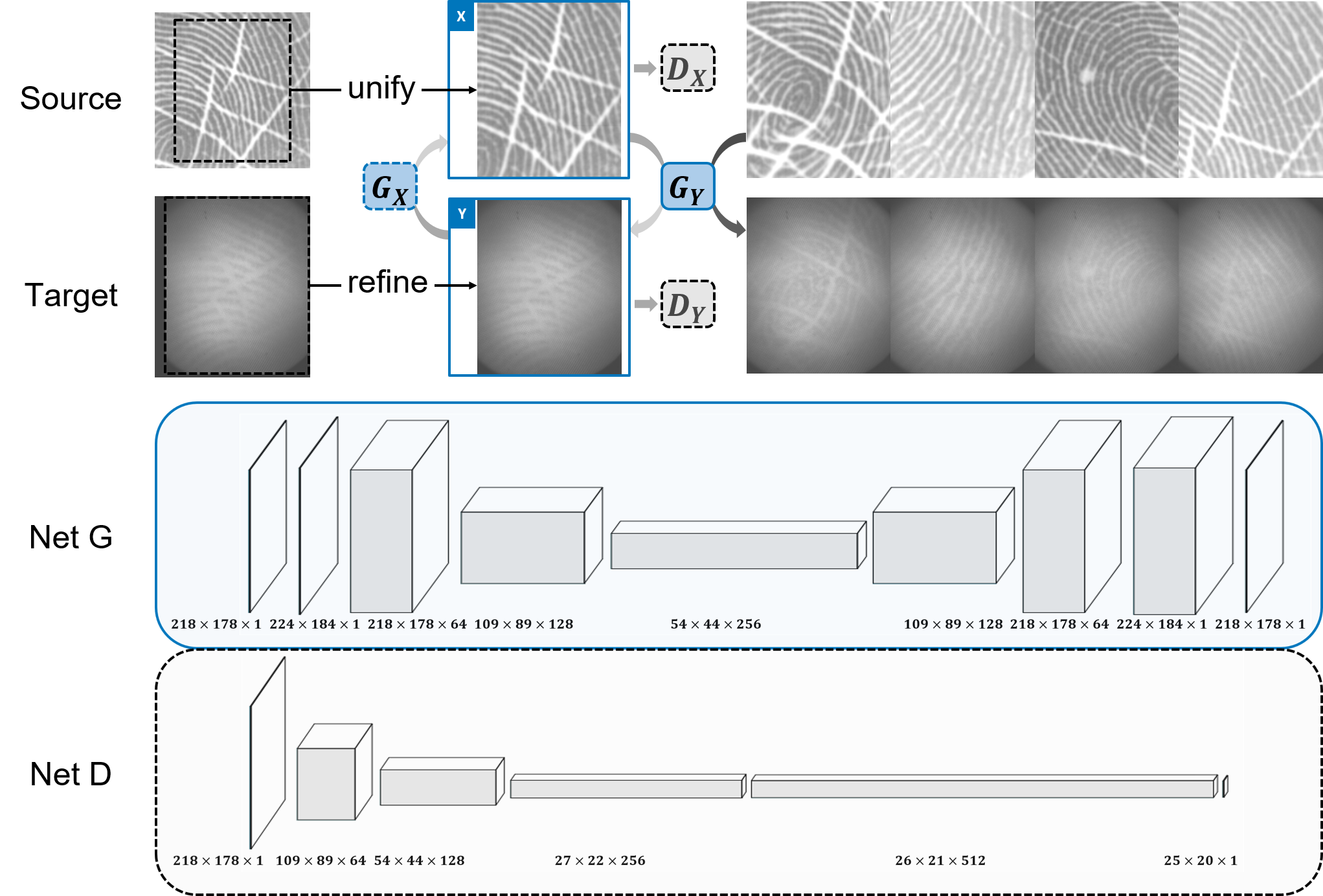}
    \caption{Style alignment from capacitive~(FPC1020) to optical in-display~(OnePlus 7P Pro).}
    \label{fig:transfer}
\end{figure}

\textbf{Experimental Setup.}
We use a capacitive touch fingerprint sensor~(FPC1020, called ``source device'') to obtain source domain images.
As shown in Figure~\ref{fig:transfer}, the image style is quite different from the target domain images captured by the optical in-display fingerprint sensor embedded on the victim device.
We use \numprint{3534} source domain images and \numprint{1507} target domain images to train a CycleGAN, and generate the fingerprint dictionary with other images from the source domain.
The raw target domain images have dark borders that hardly contain fingerprint features. We remove the border before training and get the single channel source input that are 218~px in height and 178~px in width.
The source device has a DPI 1.25 times of the victim device's fingerprint sensor and outputs 192$\times$192~px images. We unify them to coordinate with the source input.
For the network architecture in CycleGAN, we use a patch-level discriminator~\cite{isola2017image} and adopt the 9-block ResNet generator~\cite{johnson2016perceptual}. In order to make the generator outputs 218$\times$178~px images, we modify the padding of its last downsampling layer and first upsampling layer to 0. 

We use $M$~(200) matched fingerprints and $N$~(1000) unmatched fingerprints from the source domain to generate a dictionary and brute-force the victim device. In this experiment, the attack only inject images at the first sample~(i.e., triggers Error-cancel at the second FDA), and the victim device has only one fingerprint enrolled.
We measure the effect of the generated dictionary with two metrics:
(a)~$\rm{E}$: the number of times an error~(retry prompt) occurs when injecting images generated from the $N$ source images;
(b)~$\rm{M^{\prime}/M}$: the spoof~(device unlocked) rate of injecting images generated from the $M$ source images.
Note that the possible error here is not raised by failed authentication results. Instead, the authentication is terminated ahead~(most likely) by liveness detection. Therefore, the brute-force is only feasible when this error is hardly triggered.
Moreover, experiments are done to study the influence of different training epochs including zero, i.e., without style alignment.

\textbf{Results.} 
The upper right corner of Figure~\ref{fig:transfer} illustrates the effect of style alignment, where the transferred images below look much alike the refined target image in style and the fingerprint characteristics are visually reserved.
The evaluation on brute-forcing is shown in Table \ref{tab: transfer}.
The victim device raises error at each attempt if no style alignment is performed while the transferred images give no error. 
Results also show that the style alignment is the most useful with 80 epochs of training, where the time cost can be affordable and has a spoof rate at 71\%. 
In conclusion, fingerprint brute-force attack with the proposed dictionary generation method can always pass security strategies including liveness detection. More importantly, it reserves fingerprint features and have much chance to successfully spoof the SFA system.

\vspace{-1mm}

\vspace{-5mm}
\begin{table}[htbp]
\setlength{\abovecaptionskip}{0.cm}
\setlength{\belowcaptionskip}{0.1cm}
\centering
\caption{The feasibility~($\rm{E}$) and effectiveness~($\rm{M^{\prime}/M}$) experiments on brute-forcing with the dictionary generation method. The number of Epoch identifies the training epoch in the style alignment .}
\begin{threeparttable}
\begin{tabularx}{\linewidth}{lXXXXXX}
\toprule
\multirow{2}{*}{} & \multirow{2}{*}{w/o\tnote{1}} & \multicolumn{5}{c}{Epoch} \\ \cmidrule(lr){3-7}
                  &                      & 20  & 40  & 60 & 80 & 100 \\ \midrule
$\rm E$             & 1000                 & 0    & 0    & 0    & 0    & 0    \\
$\rm M^{\prime}/M$  & 0                    & 0.45 & 0.63 & 0.66 & 0.71 & 0.44   
\\ \bottomrule
\end{tabularx}
   \begin{tablenotes}
     \item[1] w/o denotes the dictionary without style alignment.
   \end{tablenotes}
   \end{threeparttable}
\label{tab: transfer}
\end{table}

\subsection{Automatic Fingerprint Brute-force Attack}

\begin{figure}[htbp]
    \centering
    \includegraphics[width=\linewidth]{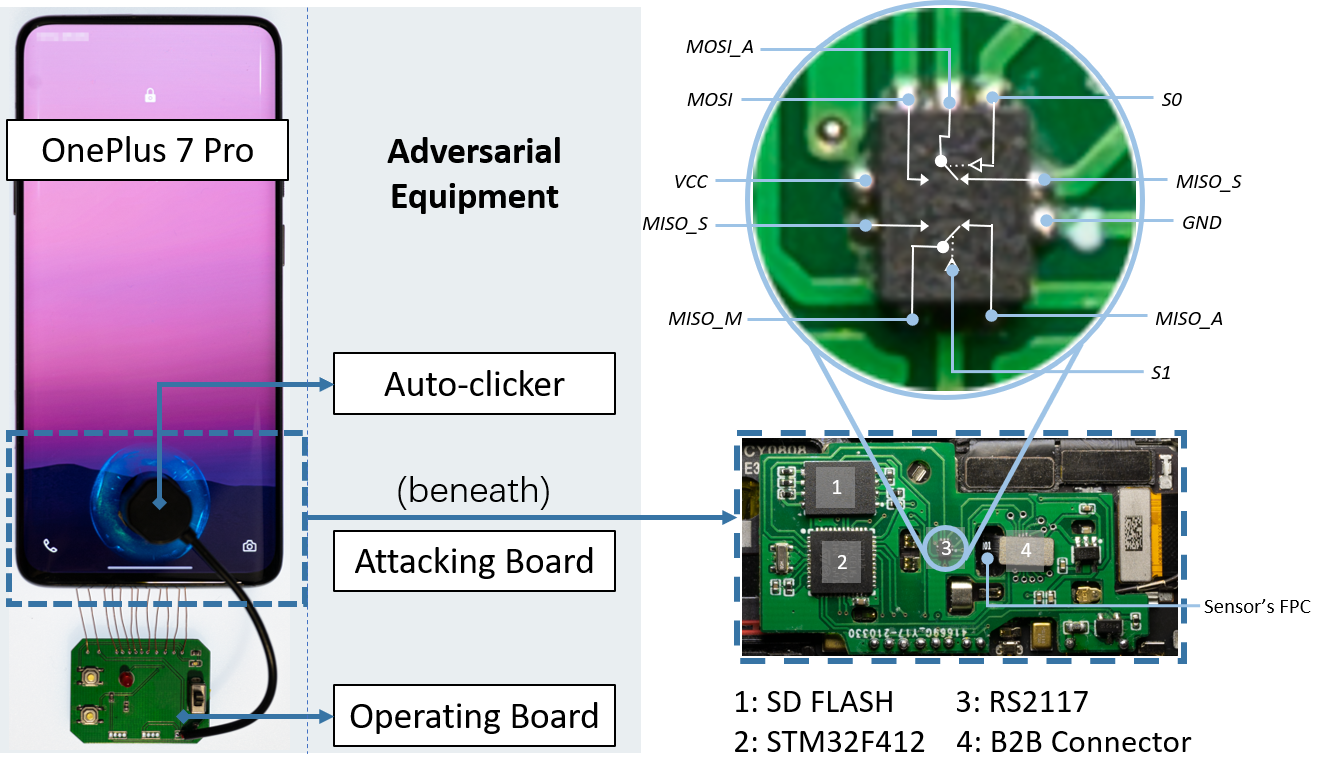}
    \caption{Example of implementing automatic fingerprint brute-force attack, which uses a suppressible attacking board, a hardware auto-clicker, and an optional operating board.}
    \label{fig:att_device}
\end{figure}

\textbf{Adversarial Equipment.}
Figure~\ref{fig:att_device} shows a snap photograph of brute-forcing the victim device. 
The attacking board, which is the core of the adversarial equipment, consists of four major components: (a)~STM32F412: the MCU that has a SPI peripheral with \textasciitilde38~MHz transmission rate in slave mode, and it controls the whole attack process, (b)~RS2117: the SPDT analog switch with 400~MHz bandwidth that switches between attack modes, (c)~SD flash: has a 8GB flash memory that can accommodate around 
\numprint{200000} 8-bit gray-scale fingerprint images, and (d)~B2B connector: connects to the smartphone motherboard and fingerprint sensor's FPC on both sides.
We also complement the adversarial equipment with an auto-clicker and an operating board to make the attack process automatic and adjustable. 
The auto-clicker is used to wake up the sensor automatically so that the large-scale attack can be simplified as injecting a pre-prepared fingerprint dictionary in the SD flash.
It works at a speed controlled by the STM32F412 MCU and make a click every second in this case to provide enough time for the FDA of each attempt.
The operating board offers a user interface for adversaries to update the fingerprint dictionary, and also support switching between collecting fingerprints and launching attacks.
The total cost of the equipment is around 15 dollars.

The implementation is made suppressible that the attacking board can be built-in. As in the figure, the connector connects to the FPC with female header on the front side and connects to the motherboard with male header on the back side.
With the suppressibility, we illustrate another attack vector that adversaries can place the attacking board secretly to steal fingerprint images.
Consider the following scenario:
the attacking board is built-in on a smartphone that lure victims to use fingerprint authentication, for example, when the smartphone is supplied by black market smartphone sellers and fluky fingerprint collection companies; those adversaries are capable of gathering a myriad of fingerprint images that are directly captured by SFA sensors.
We propose that they may use the gathered images to brute-force arbitrary devices even without style alignment.
We trial the possibility by capturing $M+N$ fingerprint images from another OnePlus 7 Pro to unlock the victim device.
Results are horrible that no error happens and the success ratio is 100\%.
Though the assumption is stronger than what we describe in Section~\ref{sec:dic_case}~(called ``zero-knowledge adversaries'' in this context), the damage is non-negligible. 

\textbf{Universal Attack Process.}
In summary, with the help of \textsc{BrutePrint}, a universal brute-forcing process for zero-knowledge adversaries only involves 4 steps: (1)~remove the rear cover of a victim smartphone and cut off the connection of fingerprint B2B to plant the adversarial equipment;
(2)~switch the operating board to the state of collecting fingerprints and make genuine attempts to collect raw fingerprint images;
(3)~generate a fingerprint dictionary with the collected raw fingerprint images and the fingerprint database that they own;
(4)~use the operating board to import the fingerprint dictionary into the SD flash and switch the state to launch automatic fingerprint brute-force attack.
The process is exactly what we show in Figure~\ref{fig:overview}. However, with an inexpensive implementation, \textsc{BrutePrint} make the inner attack logic transparent. Therefore, the threat is unprecedented that even a beginner can launch fingerprint brute-force attack without any prior knowledge about the victim.

\section{Experiments Across SFA Systems}\label{sec:experiment}

The brute-force attack is based on the infinite bypassing chances and the feasibility of fingerprint image hijacking.
In general, to bypass attempt limit, CAMF exploits the widely accepted Multi-sampling and Error-cancel mechanisms while MAL leverages some careless implementation for improving user experience.
Fingerprint image hijacking can be carried out since the fingerprint data on SPI is insecure~(e.g., unencrypted).
In this section, we broadly analyze popular SFA systems. 
We show the overall results in Section~\ref{overall}, discuss some detailed approaches and findings in Section~\ref{sec:detail}, and estimate the success rate of \textsc{BrutePrint} in Section~\ref{sec:rate}.

\subsection{Empirical Analysis}\label{overall}

\begin{table*}[htbp]
\setlength{\abovecaptionskip}{0.cm}
\setlength{\belowcaptionskip}{0.1cm}
\centering
\caption{Basic information about the 10 experimental devices. The $r_{max}$ is the maximum number of fingerprints that can be enrolled on the device. The statistics of attempt limit is given for three typical types of fingerprint applications. Some values represented with $5 \times x$ mean that a 30-second waiting period is enforced between $x$ times of 5 unsuccessful attempts.}
\begin{threeparttable}
\begin{tabularx}{\linewidth}{ccccccXXX}
\toprule
\multicolumn{4}{c}{Device}       & \multicolumn{2}{c}{Sensor}       & \multicolumn{3}{c}{Attempt Limit} \\ \cmidrule(lr){1-4} \cmidrule(lr){5-6} \cmidrule(lr){7-9}
Manuf./Model        & OS/Ver.    & TEE  & $r_{max}$    & Manuf.    & Type                 & ScreenLock\tnote{1}    & Payment\tnote{2}   & Privacy\tnote{3}   \\ \toprule
Xiaomi Mi 11 Ultra  & Android 11 & QTEE     & 5 & Goodix    & Optical (ultra-thin)\tnote{*} & 5$\times$4        & 5$\times$4        & 5         \\
Vivo X60 Pro       & Android 11 & Kinibi     & 5  & Goodix    & Optical\tnote{*}              & 5         & $\infty$         & 5         \\
OnePlus 7 Pro       & Android 11 & QTEE     & 5  & Goodix    & Optical\tnote{*}              & 5         & 5         & 5         \\
OPPO Reno Ace       & Android 10 & QTEE     & 5  & Goodix    & Optical\tnote{*}              & 5$\times$4         & 5$\times$4         & 5$\times$4         \\
Samsung Galaxy S10+ & Android 9  & Knox     & 4  & Qualcomm  & Ultrasonic\tnote{*}           & 5$\times$10        & 5         & 5$\times$10        \\

OnePlus 5T          & Android 8  & QTEE     & 5  & Goodix    & Capacitive           & 5$\times$4        & 5$\times$4        & 5$\times$4        \\ \midrule

Huawei Mate30 Pro 5G & HarmonyOS 2 & TrustedCore     & 5  & Goodix    & Optical\tnote{*}    & 5$\times$4         & 5$\times$$\infty$         & 5$\times$$\infty$         \\
Huawei P40           & HarmonyOS 2 & TrustedCore     & 5  & Novatek    & Optical\tnote{*}    & 5$\times$4         & 5$\times$$\infty$         & 5$\times$$\infty$         \\ \midrule

Apple iPhone SE     & iOS 14.5.1 & Secure Enclave     & 5  & AuthenTec & Capacitive           & 5         & 5         & 5         \\
Apple iPhone 7      & iOS 14.4.1 & Secure Enclave     & 5  & AuthenTec & Capacitive           & 5         & 5         & 5        \\ \bottomrule
\end{tabularx}
   \begin{tablenotes}
     \item[*] In-display fingerprint sensors that are incorporated under the screen.
     \item[1] A1: unlock the screen of the devices.
     \item[2] A2: make payments on pre-installed payment apps. Since OnePlus Pay is made exclusive to some countries, we use PayPal instead. For other models, the specific apps are Mi Pay, Vivo Pay, OPPO Pay, Samsung Pay, Huawei Pay, and Apple Pay.
     \item[3] A3: login pre-installed privacy protection apps. Hidden Folders for Xiaomi, Secure Folder for Samsung, File Safe for Vivo, Privete Safe for OPPO, LockBox for OnePlus, Safe for Huawei and Notes for Apple.
   \end{tablenotes}
   \end{threeparttable}
\label{tab: basic}
\end{table*}

\begin{table*}[htbp]
\setlength{\abovecaptionskip}{0.cm}
\setlength{\belowcaptionskip}{0.1cm}
\centering
\caption{Attributes related with the attacks, discovered vulnerabilities, and the influence of the attacks on the experimental devices. The examined Attributes include Samples~(the number of samples in Multi-sampling), Cancel~(the existence of Error-Cancel), Hot-Plug~(support hot-plugging or not), Decode~(can be decoded or not), $f_{SPI}$~(data transmission frequency on SPI). Three Attacks, namely Bypassing~(attempt limit bypassing), Hijacking~(fingerprint image hijacking), and Brute-force~(fingerprint brute-force attack), are tested on the three types of applications. Except Hijacking of which the results do not differ among applications, we represent the attack results with a tuple corresponding to the application A1, A2, and A3.}
\begin{threeparttable}
\begin{tabularx}{\linewidth}{lcccccccXc}
\toprule
                & \multicolumn{5}{c}{Attributes}            &          & \multicolumn{3}{c}{Attacks}    \\ \cmidrule(lr){2-6} \cmidrule(lr){8-10}
                    & Samples & Cancel & Hot-Plug & Decode & $f_{SPI}$~(MHz)       & \multirow{-2}{*}{Vulnerability}     & Bypassing & Hijacking & Brute-force       \\ \toprule
Xiaomi Mi 11 Ultra  & 2 & $\cmark$ & $\cmark$ & $\cmark$ & 32  & CAMF, MAL & ($\infty$,$\infty$,$\infty$) & $\cmark$ & ($\cmark$,$\cmark$,$\cmark$) \\
Vivo X60 Pro  & 3 & $\cmark$ & $\cmark$ & $\cmark$ & 25  & CAMF, MAL & ($\infty$,$\infty$,$\infty$) & $\cmark$ & ($\cmark$,$\cmark$,$\cmark$) \\
OnePlus 7 Pro       & 4 & $\cmark$ & $\cmark$ & $\cmark$ & 25  & CAMF & ($\infty$,$\infty$,$\infty$) & $\cmark$ & ($\cmark$,$\cmark$,$\cmark$) \\
OPPO Reno Ace  & 3 & $\cmark$ & $\cmark$ & $\cmark$ & 25  & CAMF & ($\infty$,$\infty$,$\infty$) & $\cmark$ & ($\cmark$,$\cmark$,$\cmark$) \\
Samsung Galaxy S10+ & 2$\sim$4\tnote{*} & $\cmark$ & $\cmark$ & $\cmark$ & 24  & CAMF & ($\infty$,$\infty$,$\infty$) & $\cmark$ & ($\cmark$,$\cmark$,$\cmark$) \\
OnePlus 5T          & 2 & $\cmark$ & $\cmark$ & $\cmark$ & 4.8  & CAMF & ($\infty$,$\infty$,$\infty$) & $\cmark$ & ($\cmark$,$\cmark$,$\cmark$) \\ \midrule

HUAWEI Mate30 Pro 5G & 2 & N/A\tnote{\dag} & $\cmark$ & $\cmark$ & 23 & MAL & ($\infty$,$\infty$,$\infty$) & $\cmark$ & ($\cmark$,$\cmark$,$\cmark$) \\
HUAWEI P40          & 2 & N/A\tnote{\dag} & $\cmark$ & $\cmark$ & 23  & MAL & ($\infty$,$\infty$,$\infty$) & $\cmark$ & ($\cmark$,$\cmark$,$\cmark$) \\ \midrule

Apple iPhone SE & 3         & $\cmark$     & $\cmark$ & $\xmark$ & 7.7  & CAMF & (15,15,15) &    $\xmark$       &    ($\xmark$,$\xmark$,$\xmark$)           \\
Apple iPhone 7  & 3         & $\cmark$     & $\cmark$ & $\xmark$ & 7.7  & CAMF & (15,15,15) &  $\xmark$     &  ($\xmark$,$\xmark$,$\xmark$) \\ \bottomrule
\end{tabularx}
   \begin{tablenotes}
     \item[*] The number is a range because it varies empirically, where 4 is the most frequent case.
     \item[\dag] Not triggered by the tested Error-cancel triggers.
   \end{tablenotes}
   \end{threeparttable}
\label{tab: empirical}
\end{table*}


\subsubsection{Experimental Setup}
To extend the case study to general SFA systems and different applications, \textsc{BrutePrint} addresses the following research questions:
\begin{itemize}
    \item \emph{RQ1.} For attempt limit bypassing, do CAMF and MAL vulnerabilities exist in those latest versions of popular (vendor-customized)~OSs and are they exploitable?
    \item \emph{RQ2.} For SPI MITM attack, is the hardware implementation compatible for attacking fingerprint sensors embedded on different smartphone models?
    \item \emph{RQ3.} For fingerprint data encoding and decoding, is it feasible to reverse engineer the communication protocol between sensor and processor on all smartphone models?
    \item \emph{RQ4.} Besides the ScreenLock application, can fingerprint brute-force attack be performed on high-level security required Apps~(i.e., for Payment and Privacy)?
\end{itemize}

For RQ1, CAMF is validated through the number of samples within an attempt and the effect of possible Error-cancel triggers~(i.e., checksum modification, glitch attack); MAL is examined by the feedback of fingerprint sensors in lockout mode.
For RQ2, we check the preliminary for SPI MITM that hot plugging the fingerprint sensor is supported; the adaption for different smartphone models depends on how the fingerprint sensor is connected to the processor through FPC. 
For RQ3, most of the protocols are not complicated and can be guessed as in the case study; since the protocol design relies more on the sensor~(especially its type and manufacturer), we validate the feasibility on five representatives.

The necessity of studying RQ4 is that a fingerprint gained in ScreenLock~(A1) may not pass the authentication for Payment~(A2) and Privacy~(A3) Apps.
Reasons include that: 
the matching algorithm may uses higher reference thresholds for A2 and A3; 
some Apps use dedicated fingerprint enrollment, for example, Secure Folder on the Samsung device.
The attack may also achieve different impacts on A2 and A3 since more security strategies may be adopted inside Apps.

\subsubsection{System Selection}
Experiments are done on 10 off-the-shelf smartphones, where we cover those latest OS versions and fingerprint sensor types on the market with our best effort. 
Basic information about the selected systems are given in Table~\ref{tab: basic}.
Besides the OnePlus discussed in case study, we select devices from top-5 smartphone manufacturers~\cite{canalys2021Q2}, i.e., Samsung, Xiaomi, Apple, OPPO, and Vivo.
The specific models are chosen according to their fingerprint sensors.
For the scanning technology of SFA sensors, the pioneer is capacitive scanning. A representative is Touch ID, for which we use two Apple models. To complement, we also pick an Android model equipped with capacitive sensor. 
The increasing applications of in-display techniques is the latest trend, where Goodix dominates the market and adopts optical scanning. 
We choose Mi 11 Ultra since it represents the most current advance in ultra-thin optical sensor.
A small branch of in-display sensors adopt ultrasonic scanning developed by Qualcomm, for which we choose a Samsung Glaxy S10+.
Two models from Huawei are also selected since they use the recently released HarmonyOS and embed fingerprint sensors supplied by different vendors.

\textbf{Values of Attempt Limit.} As shown in the table, limiting 5 fingerprint attempts in a period is common on off-the-shelf smartphones. Many applications also follow Google's biometric framework to implement $x$ times of the \textit{5-attempt period} before permanently lock the fingerprint authentication, where a \textit{30-second waiting period} of temporary lockout is enforced between each of them.
In the following paper, we reference AOSP to represent the temporary lockout and the permanent lockout with \texttt{LOCKOUT\_TIMED} and \texttt{LOCKOUT\_PERMANENT}, respectively.





\subsubsection{Experimental Results}
The overall results are given in Table~\ref{tab: empirical}, showing that each targeted SFA system has at least one vulnerability.
Except for iPhone, fingerprint brute-force attack is feasible on all the tested applications on all the smartphones.
Since the feasibility is based on infinite attempt limit bypassing and successful fingerprint image hijacking,
we discuss the two attacks as follows.

\textbf{Attempt Limit Bypassing.}
CAMF Vulnerability is found on almost all devices while MAL affects four specific smartphone models.
All the Android devices are exposed to unlimited fingerprint attempts with our CAMF exploitation. Specifically, they have Multi-sampling mechanism that acquires $\geq2$ fingerprint images within one attempt, and Error-cancel is validated with the triggers. On the iOS devices, CAMF exploitation is used to bypass the attempt limit of 5 to allow ultimately 15 attempts. We infer that additional strategies related with Error-cancel prevent us from unlimited bypassing. 
For HarmonyOS, we are not sure if Error-cancel holds because the tested triggers cannot bring a supposed cancellation. 
However, we make more than 90 additional attempts by leveraging MAL in \texttt{LOCKOUT\_TIMED}. With another improper implementation related with passcode lockout, we further extend the numbers to infinity. 

\begin{figure}[htbp]
    \centering
    \begin{subfigure}[b]{.2145\linewidth}
      \centering
      \includegraphics[width=.95\linewidth]{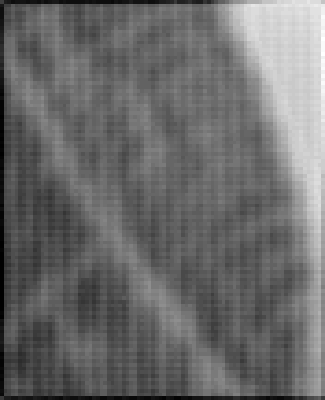}  
      \caption{Capacitive}
      \label{fig:original}
    \end{subfigure}
    \begin{subfigure}[b]{.22572\linewidth}
      \centering
      \includegraphics[width=.95\linewidth]{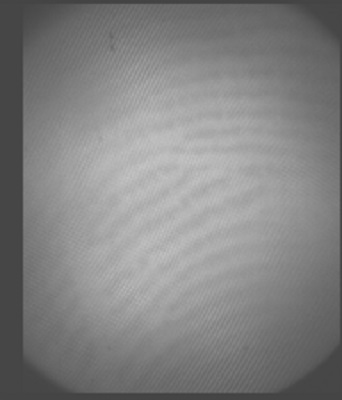}  
      \caption{Optical}
      \label{fig:original}
    \end{subfigure}
    \begin{subfigure}[b]{.27192\linewidth}
      \centering
      \includegraphics[width=.95\linewidth]{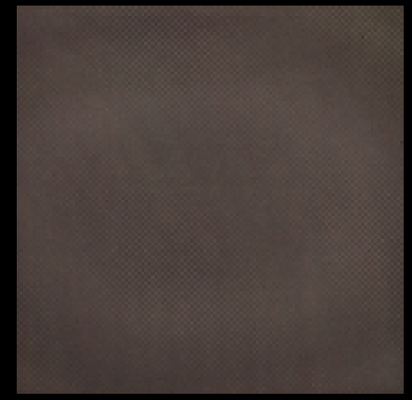}  
      \caption{Ultra-thin}
      \label{fig:original}
    \end{subfigure}
    \begin{subfigure}[b]{.23905\linewidth}
      \centering
      \includegraphics[width=.95\linewidth]{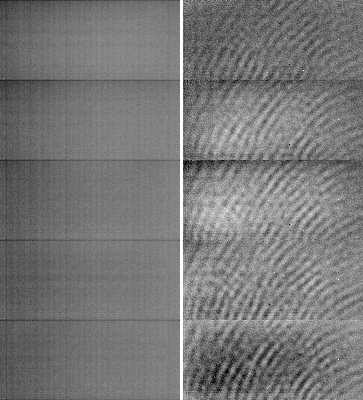}  
      \caption{Ultrasonic}
      \label{fig:original}
    \end{subfigure}
    \caption{Hijacked images for different types of fingerprint sensor. The example smartphone model is OnePlus 5T, OnePlus 7 Pro, Xiaomi Mi 11 Ultra, Samsung Galaxy S10+, respectively.}
    \label{fig:hijack_images}
\end{figure}

\textbf{Fingerprint Image Hijacking.}
Fingerprint image hijacking is feasible on all devices except for Apple, which is the only one that encrypts fingerprint data on SPI.
The hardware implementation used in case study is compatible for other smartphone models. First, we found all the smartphone fingerprint sensors have hot-plug capabilities. Second, their SPI buses all adopt the four-wire design with single master and single slave, where the transmission speeds are less than 38~MHz.
Note that the suppressibility is unnecessary in the universal attack scenario, so connecting the attacking board to the sensor with an extension FPC would be adoptable on all devices.
A slightly different connection found on the Samsung device is that the FPCs of touch screen and fingerprint sensor are integrated together, but the adversarial equipment still works as both the screen and the sensor go back to normal after hot plugging.
As shown in Figure~\ref{fig:hijack_images}, for each type of fingerprint sensor, we successfully get the fingerprint image reverse engineered from the captured data.
Moreover, as both the Mate 30 Pro and the OnePlus 7 Pro use fingerprint sensors provided by Goodix, we find the protocols share much in common. 
Therefore, with a handful of smartphone fingerprint sensor manufacturers on the market, we believe there won't be much work for the hijacking on untested smartphone models.

\subsection{Details and Findings}\label{sec:detail}

\subsubsection{\textbf{MAL-based Bypassing}}

The Xiaomi, Vivo, and Huawei devices are validated to have MAL vulnerability in A1.
As the Huawei devices only gets MAL, we take them for example to illustrate the MAL-based bypassing.

The inference attempt is unlimited essentially but cannot get chance to unlock the screen when fingerprint authentication is in \texttt{LOCKOUT\_PERMANENT}. 
In this context, we call the inference attempt made in \texttt{LOCKOUT\_TIMED} as valid since a discovered matched fingerprint can be reused in the next 5-attempt period to unlock the screen. 
We keep making attempts in \texttt{LOCKOUT\_TIMED} and launch a side-channel attack on SPI to infer the authentication results.
The side-channel information is related with the number of samples within an attempt. The fingerprint sensor acquires $2$ samples unless the fingerprint matches and stops at the first acquisition. 
Assuming that one attempt is made per second, we can make $30\times(x-1)=90$ valid inference attempts.
The number would be larger if the attempt rate is increased.
Therefore, the number of available attempts is enlarged from $20$ to more than $110$.
Note that since Hijacking is feasible on the two devices, the 110 available attempts means that $220$ fingerprint images can be submitted.

\subsubsection{\textbf{Bypassing Touch ID}}

As Touch ID encrypts the data on SPI and use a more secure TEE implementation~(i.e., Secure Enclave), we fail to achieve fingerprint image hijacking on them. However, we speculate the Error-cancel mechanism in the closed source iOS with glitch attack. The proposed CAMF exploitation compromise the attempt limit of 5 and make the actual number of available attempts 15.

\begin{table}[hbtp]
\setlength{\abovecaptionskip}{0.cm}
\setlength{\belowcaptionskip}{0.1cm}
    \centering
    \caption{The UI response (be unlocked, no retry prompt) for an attempt on Touch ID. Different screen status includes \textit{Sleep} and \textit{Wake}. Different modes for each attempt includes pressing with a enrolled ({\color{teal}{\square}}) or unenrolled ({\color{red}{\square}}) finger under \textit{Normal} circumstances and our \textit{Glitch} attack.}
    \begin{threeparttable}[b]
\begin{tabularx}{\linewidth}{ccXXXX}
    \toprule
\multicolumn{2}{c}{} & \multicolumn{2}{c}{Sleep} & \multicolumn{2}{c}{Wake} \\ \midrule
\multirow{2}{*}{Normal} & \color{teal}{\square} & $\cmark\tnote{*}$  & $\cmark$  & $\cmark$  & $\cmark$  \\  \cmidrule(lr){3-4} \cmidrule(lr){5-6}
                        & \color{red}{\square} & $\xmark$  & $\xmark\tnote{*}$ & $\xmark$ & $\xmark$ \\  \midrule
\multirow{2}{*}{Glitch} & \color{teal}{\square} & $\cmark\tnote{*}$  & $\cmark$  & $\cmark$ & $\cmark$  \\   \cmidrule(lr){3-4} \cmidrule(lr){5-6}
                        & \color{red}{\square} & $\xmark$ & $\cmark$  & $\xmark$ & $\xmark$ \\
    \bottomrule
\end{tabularx}
   \begin{tablenotes}
     \item[*] The UI responses display when less than 3 consecutive CAMF attempts happen before the current attempt.
   \end{tablenotes}
   \end{threeparttable}
    \label{tab: touchid_first}
\end{table}

\textbf{Sniff a Cancelable State.}
We firstly check whether the Error-cancel mechanism exists in iOS.
To trigger possible cancellations, we consider the glitch attack discussed in Section~\ref{sec:hijack_method}.
Figure~\ref{fig:glitch_tou} in Appendix~\ref{appendix} is an example of a successfully cancelled attempt.
That is, we carefully inspects the MISO signal and apply disturbance at the moments when fingerprint data of a sample is being transmitted.
Touch ID acquires 3 samples at most, where we find an attempt is only cancelable when the glitch is injected in 2 out of them.

We perform black-box testing and inspect the feedback on the User Interface~(UI) of iPhone.
As shown in Table~\ref{tab: touchid_first}, the inconsistency appears when we authenticate with an unenrolled finger on the Sleep device in Glitch mode, i.e., no retry prompt displays while the device is still locked.
As all other situations in Glitch mode give the same feedback as Normal, we can infer that this inconsistency is a case of Error-cancel. 
Thus, we call the condition \textit{(Glitch, Sleep)} where a failed attempt gets no UI response a \textit{cancelable state}.

\textbf{Exploit CAMF.}
For CAMF exploitation, we further analyze the cancelable state in order to fulfill two requirements:

\begin{itemize}
    \item The count of failed attempts is increased if and only if a retry prompt displays.
    The methodology is to count the number of attempts between a successful one and the last failed one that triggers passcode requirement, where a CAMF attempt is performed in the middle. Results stably show that 6 attempts with an unregistered finger can be made before the passcode requirement.
    \item Matched fingerprints can still unlock the device after CAMF attempts. 
    We validate this by authenticating with the enrolled finger after certain sequences of CAMF attempts and failed attempts. Experiments suggest that more than 2 consecutive CAMF attempts would lead to a \emph{lazy status} where any following authentication process gives no response, especially for the Sleep screen condition. For the Wake screen condition, the authentication typically reopens after 1\textasciitilde3 finger-pressings, and the status can be reset once a Normal attempt occurs.  
\end{itemize}

\noindent Therefore, we can infer that Touch ID is equipped with a counter relevant to Error-cancel, which may possibly serve to prevent ghost touches or minor hardware failures.
We successfully make 15 actual attempts, which is three times as much as the attempt limit supposes.
The whole process is described in two sub-steps as (a)~repeat the CAMF attempt twice when the screen is Sleep; (b)~make Normal attempts till a UI response is given when the screen is Wake. A Success authentication result in any of the two steps can break the process. The second sub-step is spent to ensure that next matched fingerprint can unlock the device, where adversaries can repeat from the first step until reaching the attempt limit.

\subsubsection{\textbf{Shared Fail Counter}}

We give a finding that the counter of failed fingerprint attempts is shared among applications for each user. The trace can be found in AOSP as in Listing~\ref{lst:share}, where the integer array \texttt{mFailedAttempts} and the Boolean array \texttt{mTimedLockoutCleared} are updated for an \texttt{userId}.
The issue can be leveraged to undermine the fingerprint authentication for Apps. 
Consider a scenario where an attacker has the ability to unlock the victim device, for example, when the passcode/fingerprint for screen lock~(A1) is leaked.
He may submit unlimited fingerprint attempts to gain authorization in apps by resetting the fail counter with a screen unlocking attempt.
We validate the method on all of the 10 devices: at least one of A2 and A3 can be bypassed on each device except for Samsung, while many third-party apps including PayPal are bypassed on all the devices.
In addition, the finding helps with improving the speed of brute-forcing A2 and A3 on the Huawei devices.
Recall that the two Huawei devices is bypassed with MAL exploitation for A1, while the attempt is originally unlimited in A2 and A3 except for the waiting period. 
With the exploitation of the shared fail counter, the waiting period is successfully bypassed.

\vspace{1mm}
\fbox{
  \begin{minipage}[bhtp]{.455\textwidth}
    \centering
    \lstinputlisting[]{share.java}
  \end{minipage}
}
\lstinputlisting[caption=User relevant fail counter in AOSP., label={lst:share}]{blank.java}


\subsection{Success Rate of Fingerprint Brute-force Attack}\label{sec:rate}

\begin{figure}[htbp]
    \centering
    \includegraphics[width=\linewidth]{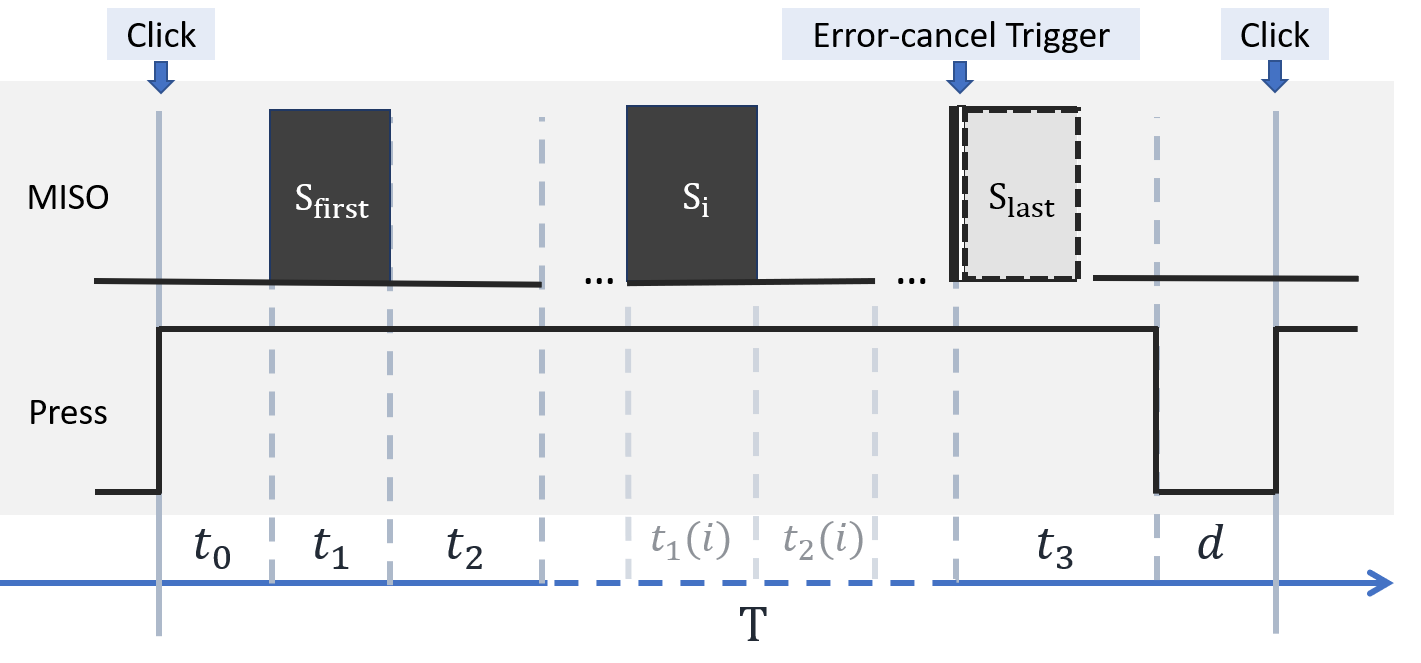}
    \caption{Time cost of one CAMF-based attempt.}
    \label{fig:time_illu}
\end{figure}

For brute-force hacking, the practicality is closely related to the actual time it takes. In particular, as most Android smartphone models observe 72-hour idle timeout before fallback to primary authentication, the fingerprint brute-force attack must take no longer than that to succeed.
For the 6 devices where the CAMF-based fingerprint brute-force attack is feasible for A1, we estimate the success rate defined in Equation~\ref{equ:suc_rate}.
The $\bm{\mathrm{FIPS}}$ in the equation is calculated with $\rm 1/T_{att}$, where $\rm T_{att}$ represents the time cost of one CAMF-based attempt.
To minimize the time, we give priority to the glitch attack in triggering Error-cancel if the fingerprint acquisition is terminated at the glitch point.
Therefore, the attempt time can be estimated as
\begin{equation}
    \mathrm{\bar T_{att}} = t_0 + (\bar{t_1} + \bar{t_2}) \cdot n + t_3 + d \,
\end{equation}
which consists of five components as in Figure~\ref{fig:time_illu}: $t_0$ is the time between finger pressing and the first transmission of fingerprint data; $t_1$ and $t_2$ sum up the time for transmitting one of the $n$ valid samples on MISO, which is spent on the fingerprint data and the time interval, respectively; $t_3$ stands for the cost between the glitch~(if exists) and the end of the pressing; $d$ is the time delay~(e.g., response time of the auto-clicker) between each attempt.

\begin{figure}[htbp]
    \centering
    \includegraphics[width=.99\linewidth]{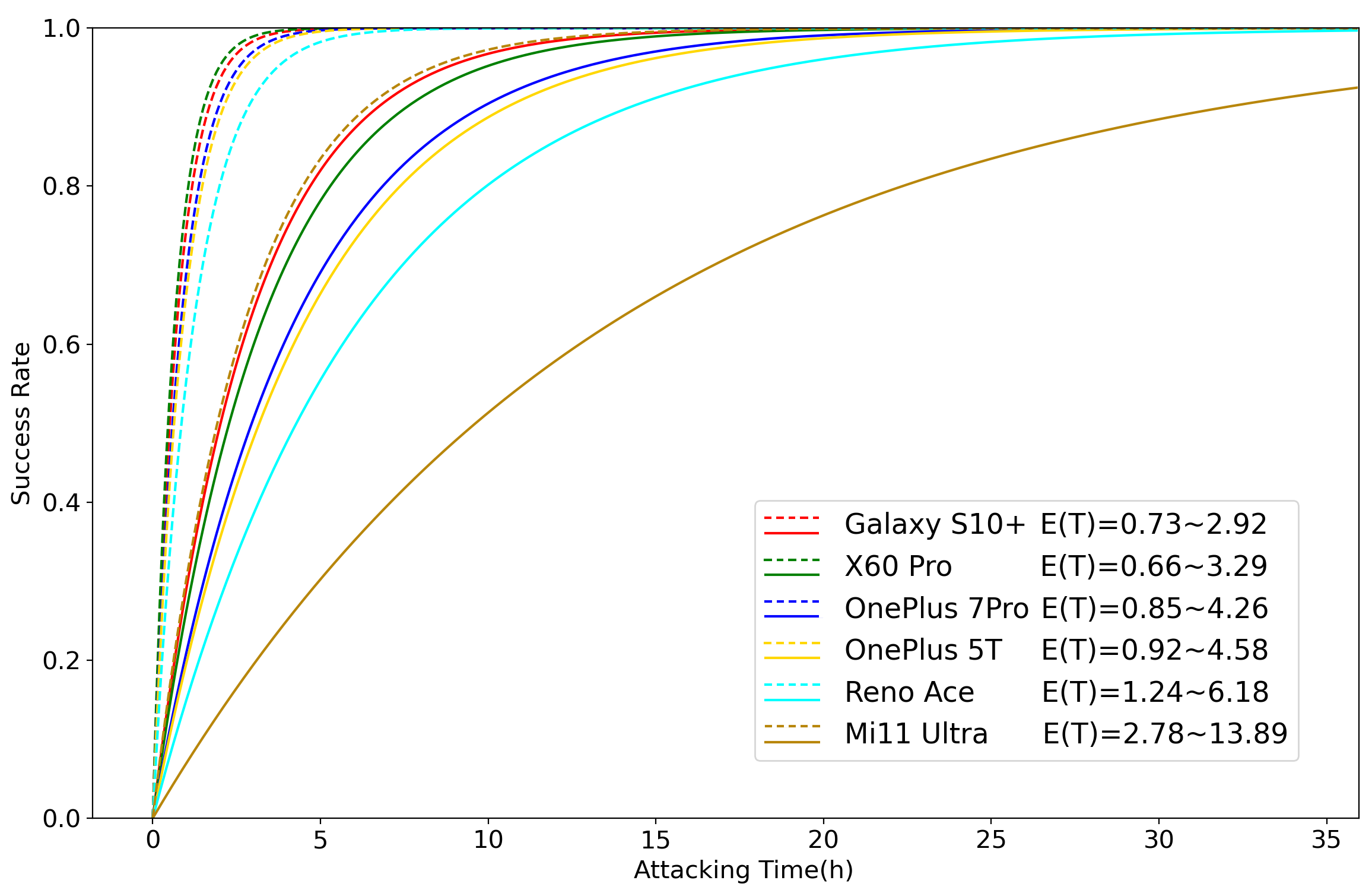}
    \caption{Success rate of fingerprint brute-force attack over time. The number of enrolled fingerprints $r$ is set to 1 for the solid lines and $r_{max}$~(see Table~\ref{tab: basic}) for the dash lines. The $\rm E(T)$ is the expected value of the time~(in hour) taken by a successful attack.}
    \label{fig:success}
\end{figure}

\textbf{Results.} 
We show the success rate over 36 hours of brute-forcing the 6 devices in Figure~\ref{fig:success}.
In the most difficult case where the victim smartphone has only one fingerprint enrolled, the expected values of the success time fall in a range of 2.9\textasciitilde13.9 hours.
Within an half of the idle timeout, nearly all the tested devices can be unlocked with an approximate 100\% success rate. The Mi11 device is the only exception where the success rate still achieves 92.51\%. 
Under the situation where the victim enrolls the maximum number of fingerprints, the speed is greatly increased and the expected time ranges from 0.66 to 2.78 hours. 
To conclude, our attack is practical and not limited by the design of the idle timeout.

\section{Mitigation}\label{sec:mitigation}

We discuss defense strategies in terms of the software framework and the hardware protocol as follows:

\textbf{CAMF Check.} To defend against the attempt limit bypassing, we propose to detect CAMF exploits by setting an additional limit for the error-cancel~(see Algorithm~\ref{CAMF_check}). For each attempt, we check whether a cancellation happens and increase the security level once the historical cancel number reaches a threshold. Specifically, the smartphone can only be unlocked with matched fingerprint images when there is no cancellation in an attempt~(Lines 8-10). 
It's worth noting that the proposed method is more secure than what we inferred from Touch ID in two ways: 1)~the cancel counter cannot be cleared with a failed attempt, 2)~all fingerprint acquisitions across a single attempt are checked even if a successful matching has occurred.
This design also balances usability with security as the unlocking time is only increased by minor cancel checks while the FRR is not influenced.

\textbf{Secure Channel.} To prevent MITM attacks, we suggest that the vendors of fingerprint sensors and smartphones are responsible to encrypt crucial data during communication. Most importantly, the pin for SPI data input, i.e., MISO, should carry high entropy messages. 
To protect from side-channel attacks, we recommend that fingerprint acquisition should behave consistently that no relevance to the matching results can be inferred.

In other aspects, it's also better to take precaution against side-channel attacks in UI level.
In fact, we find the animation of fingerprint acquisition leaks the authentication results on many smartphones, i.e., the animation time is shorter in a matched case. 
Moreover, the MAL-based bypassing teaches a lesson that the hardware implementation of actual authentication should strictly follow the software logic.

\begin{algorithm}[htbp]
\caption{Attempt limit with CAMF check}\label{CAMF_check}
$ attempt\_count \leftarrow 0$\;
$ cancel\_count \leftarrow 0$\;
\While{$attempt\_count \textless \rm{ATTEMPT\_LIMIT}$}
{
$\mathbf{F} =$ AquireFingerprints$()$\;
\If{$cancel\_count \textless \rm{CANCEL\_LIMIT}$}
    {\If{Verify$(\mathbf{F})$}{\Return{$\rm{UNLOCK}$}}}
\Else{\If{Verify$(\mathbf{F})$ \textbf{and not} CheckCancel$(\mathbf{F})$}{\Return{$\rm{UNLOCK}$}}}
\If{CheckCancel$(\mathbf{F})$}
    {$cancel\_count \leftarrow cancel\_count + 1$\;}
\Else{\tcp{Prompt ``Try Again''} $attempt\_count \leftarrow attempt\_count + 1$\;}
}
{\tcp{Prompt ``Passcode Requirement''}}
{\Return{$\rm{LOCKOUT}$}}
\end{algorithm}

\section{Discussion}\label{sec:discussion}

We discuss more concerns about the proposed attack, including the application, influence, and our future work. 

\textbf{Presentation Attack Enhancement.} Besides fingerprint brute-force attacks that require no prior knowledge about victims, the proposed method is also able to enhance traditional presentation attacks.
The first point is that attempt limit is no longer a barrier for presentation attacks on smartphones. Attackers can traverse different materials and printing techniques to find the best for making artificial replicas.
More importantly, the laborsome fabrication process can be replaced with image-level editions to some extend. The benefits are gained in two ways. 
To relieve the presentation from real objects, attackers can directly present a fingerprint latent image to the system. In other words, the attack is performed at the back-end transmission channel of the fingerprint sensor rather than the front-end.
To improve the attack performance, attackers can transform the images acquired by the fingerprint sensors.
They can apply image rotations to simply bypass the liveness detection guarded with a transformation matrix.
Another example is trying fingerprint image enhancement algorithms, e.g., Gabor filter, with various hyper-parameters to enhance the matching rate.
 
\textbf{Other Biometric Authentication Systems.}
As Multi-sampling mechanism is considered as one of the best practices in biometric authentication systems, there is reason to suspect that CAF also exists in other systems based on face, iris and palm biometric. In fact, from AOSP code, we have already found the integration of the biometric authentication framework in \codeword{android.hardware.biometrics.BiometricManager} and \codeword{android.server.biometrics.BiometricServiceBase} packages, where Error-cancel mechanism can also be seen.
To examine the vulnerability to brute-force attacks and make exploitation, some hardware-related adaption should be made as the transmission channel differs among biometric sensors. Nevertheless, the proposed mitigation measures work with all biometric systems.

\textbf{Fingerprint Dictionary Generation.}
While DeepMasterPrint~\cite{bontrager2018deepmasterprints} proposes that training a Generative Adversarial Network~(GAN) on real fingerprint image databases is possible to generate a synthetic partial fingerprint dictionary for attacks, they fail to consider the security level on off-the-shelf SFA systems. Specifically, only 1.11\% attacks are shown successful on VeriFinger~\cite{verifin} when the FAR is cut down to 0.01\%, but the criterion is required to be 0.002\% for SFA. 
Instead of generating partial fingerprint images, we find another way that use GAN to transfer the databases into styles that are acceptable for SFA systems. The method is shown effective on a real system, but there are two aspects where we aim to improve in the future. 
Firstly, the brute-forcing time relies on the fingerprint diversity of training databases, and we can reduce the time through transferring synthetic fingerprints. 
The idea from DeepMasterPrint can be borrowed to synthesize fingerprints that trigger more collisions, and we believe it's also probable to successfully attack smartphones where the attempt number is not enlarged to infinite.
Secondly, we do a case study to show the effectiveness of brute force with the transferred dictionary due to the time limit. In our next work, we plan to do large-scale experiments on the algorithm vulnerabilities across SFA systems.

\section{Related Work}\label{sec:related}

Prior works have proposed different types of presentation attacks that target traditional fingerprint authentication systems~\cite{bowden2012fooling, xue2020lopa, sousedik2014presentation}. The most widely adopted methodology is to impersonate a legitimate user via artefacts~\cite{iso2016-1}, such as latent print images~\cite{cao2016hacking} and gummy fingers~\cite{matsumoto2002impact, gonzalo2018attacking, ak17Goldfin}. 
These works differ from each other mostly on the chosen materials and printing techniques for fingerprint spoofing in front of the sensor. In comparison, our work places an attacking board in the sensor's transmission channel to inject arbitrary fingerprint images. In this way, we avoid delicate crafts and enhance the attack through effortless image edition.

On off-the-shelf SFA systems, a few presentation attacks were reported successful in industry~\cite{geekpwn2019}. For instance, the CCC team~\cite{ccc2013touchid} fabricated a latex sheet after photographing a victim's fingerprint and bypassed the iPhone 5S Touch ID. A recent study from Cisco~\cite{CT2020fincloning} selected fabric glue to make 3D printed fingers, which claimed to bypass 8 out of 13 tested devices within 20 attempts. 
In order to succeed before reaching the attempt limit, these attacks take skillful hackers laborious efforts to collect and fake victim's fingerprint.
Zhang et al.~\cite{zhang2015fingerprints} study 4 types of security pitfalls on SFA that may be exploited by malware and is the first~(in 2015) to discuss the attacks in view of the authentication framework. However, as the software security in SFA is much enhanced especially with securer TEE implementations, follow-on works hardly appear.  
Our work discovers defects in SFA frameworks that affect even the most advanced software and hardware techniques.
We make exploitation to achieve fingerprint brute-force attacks, becoming free of prior knowledge about victim's fingerprint.
 

Prior literature makes efforts to defeat presentation attacks through liveness detection~\cite{ghiani2017review, Herrero2007VulnerabilitiesIB, chugh2020fingerprint}. Hardware-based methods use additional or specialized sensors to capture biological characteristics~\cite{baldisserra2006fake, antonelli2006fake, moolla2019optical},  software-based approaches leverage features such as texture to separate live and dummy fingerprint images~\cite{chugh2018fingerprint}, and many recent works are learning-based~\cite{nogueira2016fingerprint}. 
Wu et al.~\cite{wu20liveNE} further propose to defend puppet attacks by monitoring the acceleration and rotation angle of mobile devices.
In our experiments, we bypass the liveness detection module on off-the-shelf smartphones with the help of CycleGAN. 

\section{Conclusion}\label{sec:conclusion}

This paper proposes fingerprint brute-force attacks on off-the-shelf smartphones.
We discover vulnerabilities in SFA ecosystem that the fault-tolerant mechanism / careless user-friendly implementation can be exploited through the insecure transmission on SPI to fool the unreliable authentication algorithms.
We validate the attacks on 10 representative smartphones, where all of them are affected to some extent.

With the proposed attack, adversaries can brute-force the fingerprint authentication on arbitrary victim smartphone to unlock the device and cheat many security apps. In addition, the attack method can be used to enhance presentation attacks and may also applies to other biometric systems.
The unprecedented threat needs to be settled in cooperation of both smartphone and fingerprint sensor manufacturers, while the problems can also be mitigated in OSs. We hope this work can inspire the community to improve SFA security.

\newpage






%


\appendix \label{appendix}

\subsection{Glitch Attack on Touch ID}

\begin{figure}[htbp]
    \centering
    \includegraphics[width=\linewidth]{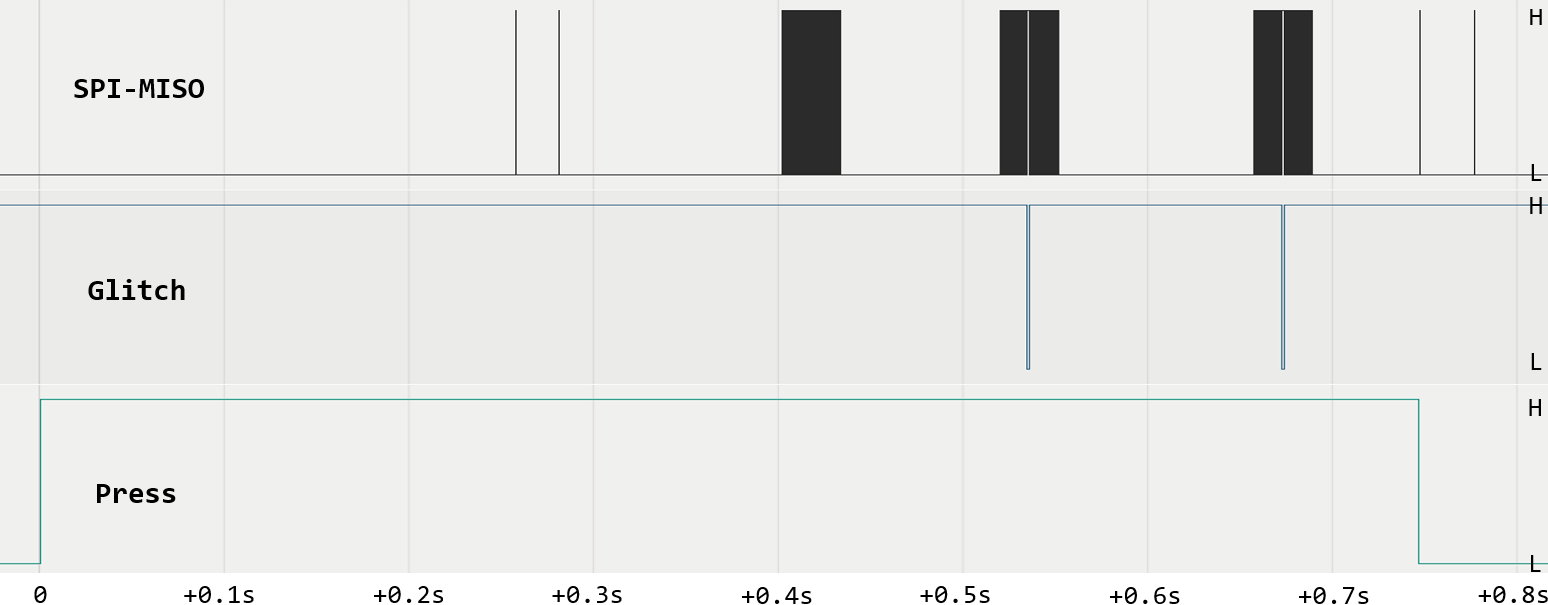}
    \caption{Signals captured during our glitch attack on Touch ID within one attempt.}
    \label{fig:glitch_tou}
\end{figure}

To trigger Error-cancel in CAMF exploitation, we use the glitch attack discussed in Section~\ref{sec:hijack_method}. Figure~\ref{fig:glitch_tou} is an CAMF attempt that successfully bypasses the attempt limit. The \textit{Glitch} signal in the figure indicates whether the connection between fingerprint sensor and processor is valid (High level) or not (Low level).

\end{document}